\documentclass{emulateapj}

\include{epsf}
\newcommand{\be}{\begin{equation}}
\newcommand{\ee}{\end{equation}}
\newcommand{\ba}{\begin{eqnarray}}
\newcommand{\ea}{\end{eqnarray}}
\newcommand{\nn}{\nonumber}
\def\simless{\mathbin{\lower 3pt\hbox
   {$\rlap{\raise 4pt\hbox{$\char'074$}}\mathchar"7218$}}}
\def\simgreat{\mathbin{\lower 3pt\hbox
   {$\rlap{\raise 4pt\hbox{$\char'076$}}\mathchar"7218$}}}
\slugcomment{2009, ApJ, 703, 285}

\shorttitle{Radio and 24 $\mu$m BLAST counterparts in CDFS}
\shortauthors{S. Dye et al.}

\begin{document}


\title{Radio and mid-infrared identification of BLAST source counterparts
in the Chandra Deep Field South}


\author{
Simon Dye\altaffilmark{1},
Peter A. R. Ade\altaffilmark{1},
James J. Bock\altaffilmark{2}, 
Edward L. Chapin\altaffilmark{3}, 
Mark J. Devlin\altaffilmark{4}, 
James S. Dunlop\altaffilmark{5},
Stephen A. Eales\altaffilmark{1},
Matthew Griffin\altaffilmark{1},
Joshua O. Gundersen\altaffilmark{6}, 
Mark Halpern\altaffilmark{3},
Peter C. Hargrave\altaffilmark{1}, 
David H. Hughes\altaffilmark{7}, 
Jeff Klein\altaffilmark{4},
Benjamin Magnelli\altaffilmark{8},
Gaelen Marsden\altaffilmark{3}, 
Philip Mauskopf\altaffilmark{1},
Lorenzo Moncelsi\altaffilmark{1},
Calvin B. Netterfield\altaffilmark{9,10}, 
Luca Olmi\altaffilmark{11,12}, 
Enzo Pascale\altaffilmark{1},
Guillaume Patanchon\altaffilmark{13}, 
Marie Rex\altaffilmark{4}, 
Douglas Scott\altaffilmark{3},
Christopher Semisch\altaffilmark{4}, 
Tom Targett\altaffilmark{3}, 
Nicholas Thomas\altaffilmark{6},
Matthew D. P. Truch\altaffilmark{4}, 
Carole Tucker\altaffilmark{1},
Gregory S. Tucker\altaffilmark{14}, 
Marco P. Viero\altaffilmark{9} \&
Donald V. Wiebe\altaffilmark{3}
}


\altaffiltext{1}{Cardiff University, School of Physics \& Astronomy, 
Queens Buildings, The Parade, Cardiff, CF24 3AA, U.K.}
\altaffiltext{2}{Jet Propulsion Laboratory, Pasadena, CA 91109-8099, 
U.S.A.}
\altaffiltext{3}{Department of Physics \& Astronomy, University of 
British Columbia, 6224 Agricultural Road, Vancouver, BC V6T 1Z1, Canada}
\altaffiltext{4}{Department of Physics \& Astronomy, University of
Pennsylvania, 209 South 33rd Street, Philadelphia, PA, 19104, U.S.A.}
\altaffiltext{5}{Institute for Astronomy, University of Edinburgh,
Royal Observatory, Edinburgh, EH9 3HJ, U.K.}
\altaffiltext{6}{Department of Physics, University of Miami, 1320 
Campo Sano Drive, Coral Gables, FL 33146, U.S.A.}
\altaffiltext{7}{Instituto Nacional de Astrof\'{i}sica \'{O}ptica 
y Electr{\'o}nica (INAOE), Aptdo. Postal 51 y 72000 Puebla, Mexico}
\altaffiltext{8}{Laboratoire AIM, CEA/DSM-CNRS-Universit\'{e} Paris Diderot, 
IRFU/Service d'Astrophysique, B\^{a}t. 709, CEA-Saclay, F-91191
Gif-sur-Yvette C\'{e}dex, France}
\altaffiltext{9}{Department of Astronomy \& Astrophysics, University of 
Toronto, 50 St. George Street Toronto, ON M5S 3H4, Canada}
\altaffiltext{10}{Department of Physics, University of Toronto, 60 St. 
George Street, Toronto, ON M5S 1A7, Canada}
\altaffiltext{11}{University of Puerto Rico, Rio Piedras Campus, Physics 
Dept., Box 23343, UPR station, Puerto Rico}
\altaffiltext{12}{INAF, Osservatorio Astrofisico di Arcetri, Largo E. 
Fermi 5, I-50125, Firenze, Italy}
\altaffiltext{13}{Universit{\'e} Paris Diderot, Laboratoire APC, 10, 
rue Alice Domon et L\'{e}onie Duquet 75205 Paris, France}
\altaffiltext{14}{Department of Physics, Brown University, 182 Hope 
Street, Providence, RI 02912, U.S.A.}


\begin{abstract}

We have identified radio and/or mid-infrared counterparts to 198 out
of 350 sources detected at $\geq 5\sigma$ over $\sim 9$ deg$^2$
centered on the Chandra Deep Field South (CDFS) by the Balloon-borne
Large Aperture Submillimeter Telescope (BLAST) at 250, 350 and 500
$\mu$m. We have matched 114 of these counterparts to optical sources
with previously derived photometric redshifts and fitted SEDs to the
BLAST fluxes and fluxes at 70 and 160 $\mu$m acquired with the Spitzer
Space Telescope. In this way, we have constrained dust temperatures,
total far-infrared/sub-millimeter luminosities and star formation
rates for each source. Our findings show that on average, the BLAST
sources lie at significantly lower redshifts and have significantly
lower rest-frame dust temperatures compared to submm sources detected
in surveys conducted at 850 $\mu$m. We demonstrate that an apparent
increase in dust temperature with redshift in our sample arises as a
result of selection effects. Finally, we provide the full
multi-wavelength catalog of $\geq 5\sigma$ BLAST sources contained
within the complete $\sim 9$ deg$^2$ survey area.

\end{abstract}





\section{Introduction}
\label{sec_intro}

\setcounter{footnote}{0}

Excluding the cosmic microwave background, approximately half of the
entire extragalactic background radiation is emitted at far infra-red
(IR) and sub-millimeter (submm) wavelengths
\citep[e.g.,][]{fixsen98,hauser01} peaking around $\sim 200\mu$m.
However, relatively little is known about the sources responsible
compared to the well explored optical Universe where studies have
enjoyed a head-start of several decades. Combining this with
significant recent advances in submm instrumentation, it is therefore
not surprising that galaxy surveys are now turning to the submm in the
search for a more complete understanding of the formation of
structure in the Universe.

The first deep submm surveys \citep[e.g.][]{smail97,hughes98,barger98}
revealed a population of highly energetic dust obscured sources.
Several clues suggested links between these systems and local
ellipticals such as their similar volume densities
\citep{scott02,dunne03} and clustering properties
\citep[e.g.,][]{almaini03,blain04} and their ability to rapidly form
large stellar populations. Despite these early advances towards
understanding the nature of the submm population, the small areal
coverage common to the early surveys yielded a low number of sources,
resulting in the usual limitations due to small number statistics and
sampling variance. This motivated the largest and last of the surveys
conducted using the Submillimeter Common User Bolometer Array
\citep[SCUBA;][]{holland99}, the SCUBA HAlf Degree Extragalactic
Survey \citep[SHADES;][]{mortier05}. SHADES detected a total of $\sim
120$ sources (down to $\sim 3.5 \sigma$) at 850 $\mu$m over an area of
$\sim 700$ arcmin$^2$ \citep{coppin06}.

Although the large homogeneous sample of submm sources detected by
SHADES significantly enhanced previous samples, the survey still has
three main deficits: 1) The areal coverage is small and therefore
highly susceptible to sampling variance compared with optical surveys,
2) The SCUBA population appears to only represent a small fraction of
the Universe's obscured star formation \cite[see for
example][]{chapman05,coppin06,dye07a}.  This is perhaps not surprising
given that the energy of the far-IR/submm background at 850 $\mu$m is
$\sim 30$ times less than at $\sim 200\mu$m where it peaks.  3) The SCUBA
population seems to reside almost exclusively at redshifts $z
\simgreat 1$, preventing proper investigation of the link between
distant dusty galaxies and the local population.

This paper is concerned with a large new submm survey centered on the
Chandra Deep Field South (CDFS) recently carried out by the
Balloon-borne Large Aperture Submillimeter Telescope
\citep[BLAST;][]{devlin04,pascale08,devlin09}.  The survey covers
$\sim 9 {\rm deg}^2$ at each of the three BLAST wavelengths of 250,
350 and 500 $\mu$m. This is a leap of nearly two orders of magnitude in
area compared with SHADES and energetically much nearer the peak of
the far-IR/submm background.

BLAST bridges the gap between the longest wavelength channels
available to the Spitzer Space Telescope ({\sl Spitzer}) at 24, 70 and
160 $\mu$m and SCUBA at 850 $\mu$m. As we show in this paper, BLAST is
substantially more sensitive to galaxies at $z \simless 1$ where SCUBA
found very few sources yet maintains an overlap with SCUBA's
sensitivity to moderate redshifts. For example, a 10 mJy 850 $\mu$m
source at $z=1$ with a dust temperature of 30 K has a typical flux of
80 mJy at 350 $\mu$m, well within the sensitivity levels reached by
BLAST.

In order to derive scientific conclusions from the data in the same
vein as for previous submm surveys, the first step is to identify
counterparts to the sources detected by BLAST.  Like all single dish
submm observations made to date, the large beam size of BLAST
precludes direct association with sources at optical to mid-IR
wavelengths. A well proven procedure is to identify counterparts to
submm sources using radio interferometry. At radio wavelengths, submm
sources out to a redshift of $\sim 3$ can be readily detected. The
typically low surface number density of radio detections results in a
very low rate of chance alignments. In addition to radio counterparts,
sources detected at 24 $\mu$m by {\sl Spitzer} also prove useful for
this purpose for the same reasons, albeit with typically lower
positional accuracy \citep[e.g.][]{ivison07,dye08}.

The main objective of this paper is to provide radio and 24 $\mu$m
identifications of counterparts to the sources detected by BLAST in
the CDFS. We have fitted spectral energy distributions (SEDs) to the
three BLAST fluxes at 250, 350 and 500 $\mu$m and fluxes obtained by
{\sl Spitzer} at 70 and 160 $\mu$m.  In this way, we have obtained
best fit rest-frame dust temperatures, bolometric luminosities and
star formation rates (SFRs). In Section \ref{sec_data} we describe the
BLAST observations and the supporting multi-wavelength data used in
our analyses. Section \ref{sec_methodology} outlines our methodology,
in particular the procedure used for identifying counterparts in the
radio and at 24 $\mu$m.  The results are presented in Section
\ref{sec_results}. Finally, Section
\ref{sec_summary} summarizes our findings and briefly discusses the
implications of our results.

The multi-wavelength catalog of $\geq 5\sigma$ BLAST sources is
given in the appendix in Table \ref{tab_blast_cat} and contains
350 sources.

Throughout this work, we have assumed the following cosmology:
$\Omega_m=0.28$, $\Omega_\Lambda=0.72$, H$_0=$70 km\,s$^{-1}$.

\section{Data}
\label{sec_data}

\begin{figure}
\epsfxsize=80mm
{\hfill
\epsfbox{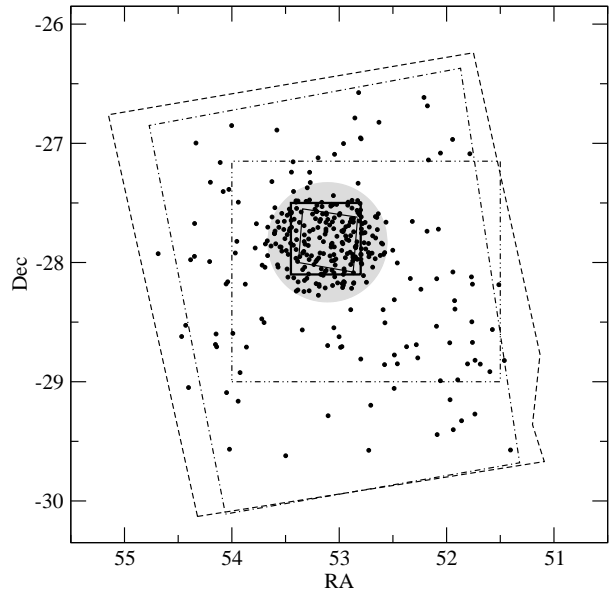}
\hfill}
\caption{Geometry of the datasets used in this work: BLAST (BGS-Wide
shown by the dashed line with BGS-Deep shaded gray), SWIRE 24 $\mu$m
(dot-dash), Norris et al.  radio catalog (dot-dot-dash), VLA CDFS
radio data (thick continuous line) and the FIDEL 24 $\mu$m catalog
(thin continuous line). The points show the $\geq 5 \sigma$ BLAST
sources.
\label{fields}}
\end{figure}

Figure \ref{fields} plots the geometry of the various datasets used in
this paper. This section gives details of those datasets and a brief
description of the BLAST dataset itself. We refer the reader to
\citet{pascale08} for a more detailed description of the primary
characteristics of BLAST and \citet{truch08,truch09} 
for information on calibration and data reduction.

\subsection{BLAST Data}
\label{sec_blast_data}

The 250, 350 and 500\,$\mu$m observations analyzed in this paper were
acquired during the BLAST 2006 flight launched on the 21st of December
from McMurdo Station, Antarctica (BLAST06).  The data encompass
the CDFS and cover 8.7\,deg$^2$ centered on ($3^{\rm h}32^{\rm m}$,
$-28^\circ 12'$) with mean $5 \sigma$ sensitivities of 180, 150 and
100\,mJy/beam at 250, 350 and 500\,$\mu$m respectively.  A deep region
of 0.8\,deg$^2$ is located within this 8.7\,deg$^2$, centered on the
southern field of the Great Observatories Origins Deep Survey
\citep[GOODS;][]{dickinson03} at ($3^{\rm h}32^{\rm m}30^{\rm s}$,
$-27^\circ48'$). This deeper region, referred to hereafter as 'BLAST
GOODS South Deep' (BGS-Deep), reach mean $5\sigma$
sensitivities of 55, 45 and 30\,mJy/beam at 250, 350 and 500\,$\mu$m
respectively \citep{devlin09}. The area surrounding BGS-Deep is
referred to hereafter as BGS-Wide.

The BLAST time series data in each of the three bands were
reduced using a custom-made pipeline \citep{pascale08}.  Periodic
observations of VY CMa conducted throughout the flight provide an
absolute calibration for the telescope gain with uncertainties of
10\%, 12\% and 13\% at 250, 350 and 500\,$\mu$m respectively
\citep[although the calibration is highly correlated --
see][]{truch08}.  Signal and variance maps with $10''\times 10''$
pixels were generated using the maximum-likelihood map-making
algorithm {\sl SANEPIC} \citep{patanchon08}.  In order to suppress
residual large-scale noise, the {\sl SANEPIC} maps were spatially
filtered to whiten structure on scales larger than the telescope field
of view ($14' \times 7'$).  A catalog of point sources with a
significance of $\geq 3 \sigma$ for each band was compiled using a
source-finding algorithm which selects the peaks in a smoothed map
produced by the inverse variance weighted convolution of the image
with the telescope point spread function \citep{devlin09}.

The source lists were synthesized into a common catalog
using a procedure which accounts for the significance and positional
uncertainty of the counterparts in each band.  The radius of the $1
\sigma$ positional error circle, $\sigma_p$, for a submm galaxy in a
catalog with signal to noise $\mu$ which has not been corrected for
Eddington bias can be expressed as 
\be
\label{eqn:searchrad} 
\sigma_p = 0.9 \, \theta \, [\mu^2 -(2\alpha +4)]^{-1/2}
\ee
for power-law counts of the form $n(>f) \propto f^{-\alpha}$, where
$\theta$ is the full width at half maximum (FWHM) of the telescope
beam \citep{ivison07}.  The BLAST06 telescope beam is best fit by
Gaussians with a FWHM of 36\,arcsec, 42\,arcsec and 60\,arcsec at 250,
350 and 500\,$\mu$m respectively \citep{pascale08}.  Using this
formula, error circles were calculated for each source in the
individual catalogs assuming the slope of
$\alpha = 2.5$ measured by \citet{devlin09}.  A minimum $1 \sigma$ 
error circle of $5''$ was imposed, equal to the $1 \sigma$ pointing
uncertainty of the maps.

The combined catalog is comprised of all sources with a significance
$\geq 5 \sigma$ in at least one band.  Sources from the other bands
were considered counterparts if they were located within twice the
radius of their respective error circles added in quadrature.
Positions of sources in the resulting combined catalog were computed
by averaging all the positions, weighted by $\sigma_p^{-2}$.  

The catalog contains 350 sources in total as listed\footnote{The BLAST
data used in this paper (including maps) are also available for
download at \url{http://blastexperiment.info}} in Table
\ref{tab_blast_cat}. The number of sources detected at $\geq 5 \sigma$
by band is 178, 145 and 168 at 250, 350 and 500 $\mu$m
respectively. 235 of the sources are located within BGS-Deep. Out of
these, the sources detected at $\geq 5
\sigma$ by band is 121, 113 and 124 at 250, 350 and 500 $\mu$m
respectively.

\subsection{Radio Data}

We have used two different catalogs for our radio counterparts. The
first contains sources detected in the 1.4GHz map observed by
\cite{miller08} using the Very Large Array (VLA).  The map covers an
area of 0.33 deg$^2$ centered on the GOODS region to an average rms
sensitivity of $\sim 8 \mu$Jy per $2.8'' \times 1.6''$ beam. The
published source catalog takes a very conservative
detection threshold of $7 \sigma$. We therefore carried out our own
source extraction on the map down to a lower detection threshold of $5
\sigma$. This results in a larger source surface number density of
$\sim 0.7$ arcmin$^{-2}$ compared to $\sim 0.4$ arcmin$^{-2}$ in the
originally published catalog. Although this increases the risk of
introducing spurious sources, our Monte Carlo method of associating
counterparts to the BLAST sources takes this into account.

For the second radio catalog, we used the shallower but wider 1.4GHz
survey of \citep{norris06} acquired using the Australia Telescope
Compact Array. This survey covers a $\sim 4$ deg$^2$ area ($2.2^\circ
\times 1.8^\circ$) centered on ($3^{\rm h}31^{\rm m}$,$-28^\circ06'$).
The rms sensitivity of the survey is 40 $\mu$Jy per $\sim 5''$ beam
giving rise to a source number density of 0.05 arcmin$^{-2}$.

We carried out our identification of radio counterparts to the submm
sources separately on each catalog. When matching to sources in the
wider Norris et al. catalog, we excluded the region covered by the
deeper VLA data. We verified that all of the Norris et al. sources in
this region were contained within the VLA data.

\subsection{24 $\mu$m Data}

Similar to the radio data, our list of 24 $\mu$m {\sl Spitzer}
counterparts comprises two separate catalogs. The first is taken from
\citet{magnelli09} as part of the Far-Infrared Deep Extragalactic
Legacy Survey \citep[FIDEL;][]{dickinson07}. The data cover 0.23
deg$^2$ ($30' \times 27'$) centered on the GOODS region.  The $5
\sigma$ point source sensitivity of FIDEL is $\sim 30 \mu$Jy giving a
source surface number density of 12 arcmin$^{-2}$.

The second 24 $\mu$m {\sl Spitzer} catalog was taken from the Spitzer
Wide-area InfraRed Extragalactic survey \citep[SWIRE;][]{lonsdale04}
third data release\footnote{See
http://irsa.ipac.caltech.edu/data/SPITZER/SWIRE}. The 24 $\mu$m SWIRE
data in the CDFS cover $\sim 8$ deg$^2$ ($2.4^\circ \times 3.4^\circ$)
centered on ($3^{\rm h}32^{\rm m}$,$-28^\circ15'$). The average
$5 \sigma$ point source sensitivity is $\sim 100 \mu$Jy although
only 24 $\mu$m sources detected at $\geq 15 \sigma$ are 
included in the catalog.

In the same manner as our identification of radio counterparts, we
identified 24 $\mu$m sources separately for both 24 $\mu$m catalogs
excluding the deeper FIDEL region when searching for SWIRE
counterparts. We verified that all 24 $\mu$m SWIRE sources are
contained within the FIDEL data.

\subsection{70 \& 160 $\mu$m {\sl Spitzer} Maps}

In addition to the three BLAST fluxes, we also extracted flux from the
SWIRE 70 and 160 $\mu$m {\sl Spitzer} maps to improve constraints on
our SED fitting. Both maps cover a rectangular area of $\sim 2.3^\circ
\times 3.5^\circ$ centered on the 24 $\mu$m catalog ($3^{\rm h}32^{\rm
m}$,$-28^\circ15'$). All bar five of the $\geq 5\sigma$ BLAST
sources with identified radio and/or 24 $\mu$m counterparts are located
within these maps.

\subsection{Photometric redshifts}

For the purpose of fitting SEDs to the BLAST and {\sl Spitzer}
photometry, we matched identified counterparts to optical sources with
previously estimated photometric redshifts. We used the two main
photometric redshift catalogs publicly available in CDFS:

\begin{itemize}

\item The photometric redshift catalog of \citet{wolf04} from COMBO-17
(Classifying Objects by Medium-Band Observations in 17 filters) covers
an area of 0.25 deg$^2$ ($0.5^\circ \times 0.5^\circ$) centered on
GOODS. Redshifts are derived from optical photometry in five broad
bands ($U$, $B$, $V$, $R$, $I$) and 12 interspersed medium bands. The
surface number density of sources is 67 arcmin$^{-2}$. Redshifts
extend up to $z \simeq 2$ and have a median of $\sim 0.9$.

\item The photometric redshift catalog of \cite{mrr08} covers a total
area of $\sim 5$ deg$^2$ ($2.2^\circ \times 2.3^\circ$ but with
gaps) centered on ($3^{\rm h}32^{\rm m}$,$-28^\circ30'$). The surface
number density of sources is $\sim 9$ arcmin$^{-2}$. Redshifts are
estimated from SWIRE photometry: The five broad optical bands u, g,
r, i and z and the two IRAC channels 3.6 and 4.5 $\mu$m. The median
redshift is $\sim 0.9$ with 10\% of galaxies at $z>2$ and 4\% at
$z>3$.

\end{itemize}

Table \ref{tab_src_props} lists the COMBO-17 and/or SWIRE
photometric redshifts matched to the counterparts where available.
Figure \ref{z_cf} shows the comparison for counterparts where
both COMBO-17 and SWIRE redshifts exist.

\section{Methodology}
\label{sec_methodology}

\subsection{Identification procedure}
\label{sec_id_proc}

To identify radio and 24 $\mu$m counterparts, we applied the
frequentist technique of \cite{lilly99} based on the method of
\cite{downes86}. The method searches for objects close to the submm
source and estimates the probability of each object being a chance
alignment using a Monte Carlo approach. We applied the approach to the
BLAST multi-wavelength source list described in Section
\ref{sec_blast_data} in the following way:
\begin{itemize}

\item[1)] Select a random position within the area common to the
BLAST and radio catalogs.

\item[2)] Find the minimum of the quantity $S=r_{sep}^2 n(>f)$ for
each radio source within a separation cutoff of $20''$ of the random
position (see below for justification of this cutoff). Here, $r_{sep}$
is the radial separation between the radio source and the random
position and $n(>f)$ is the surface number density of radio sources
brighter than the radio source.

\item[3)] Repeat steps 1) and 2) for $N$ realizations to determine
the distribution of $S$ for the radio sources.

\item[4)] Repeat steps 1) - 3) with the 24 $\mu$m catalog to determine
the distribution of $S$ for the 24 $\mu$m sources.

\end{itemize}

\begin{figure}
\epsfxsize=80mm
{\hfill
\epsfbox{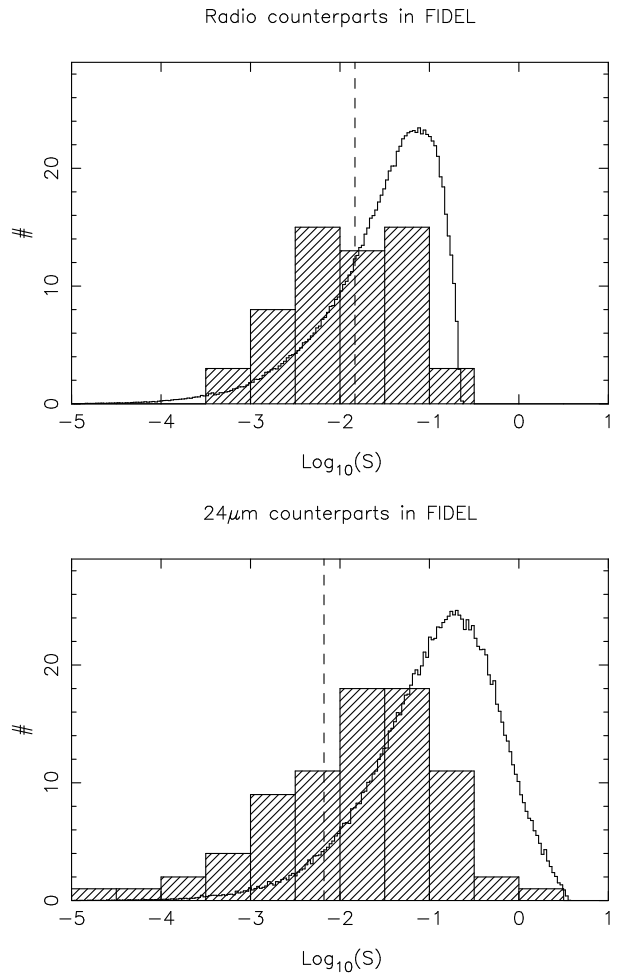}
\hfill}
\caption{The distribution of log$_{10}$($S$) for the radio and
24 $\mu$m data in the FIDEL region. The shaded histogram in each plot
shows the distribution for the most likely counterparts lying within
the separation cutoff of $20''$. The open histogram shows the
distribution of lg($S$) for one million randomly placed points (see
text). The fact that the shaded histograms are skewed to lower $S$
with respect to the open histograms indicates that a subset of the
counterparts identified in the radio and at 24 $\mu$m genuinely trace
the BLAST sources and are not completely unassociated (see Table
\ref{tab_src_props}). The dashed line corresponds to the threshold
$P=0.05$. (Note that this does not constrain 5\% of the open
histogram.  Computation of $P$ includes the random positions that do
not contain any radio/24 $\mu$m sources which are not included in the
histogram.)
\label{s_dist}}
\end{figure}

The distribution of $S$, $D(S)$, then allows the probability $P(<S_i)$ 
to be computed from a value $S_i$ of a real potential counterpart $i$:
\be
P(<S_i) = \frac{1}{N} \int^{S_i}_{0} D(S)dS \, .
\ee
Note that, in general, up to a critical surface number density,
$\int^{\infty}_{0} D\,dS<N$ since a certain fraction of the randomly
chosen positions will not contain any radio (or 24 $\mu$m) sources
within $20''$. Figure \ref{s_dist} shows the distributions obtained
for radio and 24 $\mu$m counterparts in the FIDEL region.

The most likely counterpart is that with the lowest value of $P$.
Introducing a threshold in $P$ gives a criterion for establishing
BLAST sources without counterparts.  In the analysis that follows
later in this paper, we have only included counterparts with
$P\leq0.05$. Sources with radio and 24 $\mu$m counterparts were
included if either or both counterparts satisfy $P\leq0.05$.  Although
not included in any analysis, we have also listed counterparts in the
area outside the FIDEL region with $0.05 < P\leq 0.1$. As we discuss
in Section \ref{sec_IDs}, this is justified on the basis that relaxing
the threshold to 0.1 results in another 29 counterparts at the mere
cost of including an expected two additional false counterparts.

\subsubsection{Determination of the separation cutoff}

Selection of the separation cutoff depends on several factors. Clearly
a smaller cutoff gives rise to fewer counterparts, increasing the
likelihood of missing a true counterpart. Conversely, a larger
cutoff increases the risk of associating a very bright but unrelated
counterpart (i.e., $S$ is small since $n(>f)$ is small even though
$r_{sep}$ is large). An added complication is that larger cutoffs
increase the probability of overlapping positional error circles
thereby complicating the matching process.

The separation cutoff should be a compromise between these factors.
We have therefore selected our cutoff as the radius where the expected
number of excluded true counterparts is equal to the expected number
of false counterparts. The procedure we used to determine this is as
follows:
\begin{itemize}

\item Using a large separation cutoff (e.g., $30''$), perform an 
initial ID analysis as outlined above to compute values of $P$ for all
counterparts within the cutoff. Form the distribution of radial
offsets of the primary counterparts (i.e., those counterparts with the
lowest $P$ per BLAST source).

\item Compute the number of expected false IDs by summing $P$ over
all primary counterparts. Repeat the ID analysis with decreasing
values of the separation cutoff, computing the expected number of
false IDs each time. This gives the false ID rate, i.e., the number of
false IDs per interval separation as a function of the separation
cutoff.

\item Subtract the false ID rate from the distribution of radial
offsets determined in the first step. The result is the expected
distribution of true counterparts which should have the form
$r\,e^{-r^2/2\sigma^2}$ \citep[e.g.,][]{ivison07}. Fit this to derive
$\sigma$ and hence determine the number of true counterparts that
would be excluded as a function of separation cutoff.

\end{itemize}

\begin{figure}
\epsfxsize=80mm
{\hfill
\epsfbox{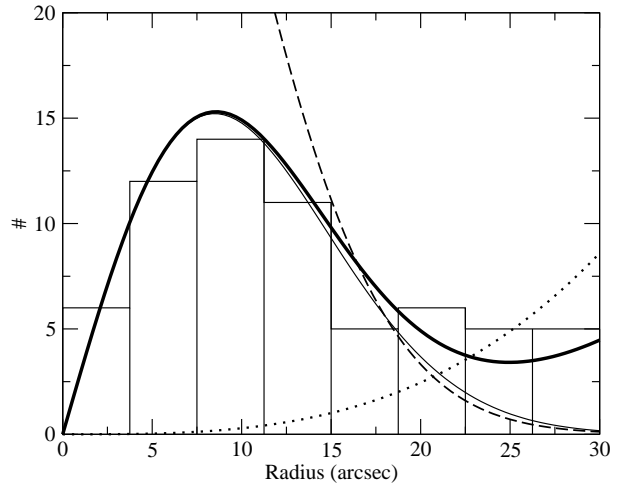}
\hfill}
\caption{Distribution of radial offsets between BLAST sources and
radio counterparts in the FIDEL region (open histogram).  The dotted
line shows the expected cumulative number of false radio IDs in this region
as a function of the separation cutoff used in the
matching. Subtracting the false ID rate (proportional to the gradient
of the dotted line) from the histogram and fitting the result with a
curve of the form $r\,e^{-r^2/2\sigma^2}$ gives the thin continuous
line with $\sigma\simeq 8''$. The sum of this curve and the false ID
rate gives the heavy continuous line which follows the histogram of
measured radial offsets. Finally, the dashed curve shows the cumulative
number of true radio counterparts that would be excluded at the given
separation cutoff. The optimal separation cutoff is selected as the
radius where the expected number of false IDs is equal to the expected
number of excluded true counterparts. In this case, for the radio
counterparts in FIDEL, the optimal cutoff is $21''$.
\label{offsets_radio_fidel}}
\end{figure}

Figure \ref{offsets_radio_fidel} shows the distribution of radial
offsets of the primary radio counterparts within $30''$ of each BLAST
source in the FIDEL region. The figure shows that the expected number
of false counterparts is equal to the expected number of excluded true
counterparts at a separation cutoff of $21''$. Repeating this analysis
for the 24 $\mu$m counterparts in the FIDEL region, we found an
optimal separation cutoff of $19''$. Similarly, the procedure returns
optimal separation cutoffs of $28''$ and $26''$ for the radio and 24
$\mu$m counterparts in the region outside of FIDEL respectively. With
this in mind, we chose a value of $20''$ as the separation cutoff for
the FIDEL region and $25''$ for the region outside FIDEL.  Whilst this
is slightly conservative for the outer-FIDEL region, the fraction of
real counterparts we would expect to exclude is low.  Similarly,
although this cutoff is slightly larger than the optimal value for the
24 $\mu$m counterparts within the FIDEL region, the false ID rate is
sufficiently low around $20''$ to make a negligible increase in the
number of expected false 24 $\mu$m counterparts. Section \ref{sec_IDs}
quantifies the resulting expected number of false IDs for each
counterpart catalog and region.

Note that the distribution of radial offsets in Figure
\ref{offsets_radio_fidel} indicates that 68\% of radio counterparts
are within an offset of $\sim 12''$. The position error of a $5\sigma$
BLAST source detected at 250 $\mu$m from equation \ref{eqn:searchrad}
is $\sim 8''$. These values are consistent once positional
uncertainties in the radio catalog and pointing errors on both the
BLAST and radio catalogs are accounted for.

\subsection{Matching to the photometric catalogs}

Prior to matching to the photometric redshift catalogs, the
identified radio counterparts were position-matched to the 24 $\mu$m
counterparts. We used a separation tolerance of $3''$ for this
matching based on the 24 $\mu$m data point spread function (PSF) of
FWHM $\simeq 6''$ and the average radio data PSF of FWHM $\simeq 5''$.
In cases where a BLAST source has a radio and a 24 $\mu$m counterpart
that both satisfy the selection criteria but are separated from each
other by more than $3''$, we selected the counterpart with the lowest
value of $P$.

All identified BLAST source counterparts were then position-matched to
the photometric redshift catalogs. For BLAST sources with a radio and
a 24 $\mu$m counterpart, the average of the radio and 24 $\mu$m
co-ordinates were used in the matching (since the radio and 24 $\mu$m
PSFs are similar). We used the same positional tolerance of $3''$ in
matching up with the redshift catalogs. We found that in all cases
where a counterpart is matched by both a COMBO-17 and SWIRE redshift
(i.e., they are both within $3''$ of the counterpart), the COMBO-17
and SWIRE sources are within $1''$ of each other.

\subsection{SED fitting}
\label{sec_sed_fitting}

We determined 250, 350 and 500 $\mu$m BLAST fluxes and the 70 and 160
$\mu$m {\sl Spitzer} fluxes for the BLAST sources with identified
photometric redshifts by reading off the flux in the beam-convolved
maps at the position of the radio/24 $\mu$m counterpart.  These fluxes
were then fitted with a modified black body SED of the form
$A\nu^{\beta}{\rm B}(\nu,{\rm T})$, where $A$ is a normalization
constant, B is the Planck function and $\beta$ was
fixed\footnote{Allowing $\beta$ to vary in the minimization generally
results in an improvement in $\chi^2$ of less than 1, implying that
the data do not support the introduction of an extra parameter.  This
was also found to be true when a secondary dust component was added.}
at the value of 1.5. 

In fitting the SEDs, we assumed a uniform prior for the temperature
over 10 K$\leq T \leq $50 K and fixed the redshift at the photometric
redshift of the optical counterpart. In cases where a redshift exists
from both COMBO-17 and the catalog of \citet{mrr08}, we selected the
redshift that minimizes $\chi^2$ which was computed allowing for
the BLAST data covariance \citep[see][]{truch08}:
\ba
\chi^2=&\,& \sum_{i,j}
(f_i-f^m_i(\beta,T,z))\sigma^{-1}_{ij} (f_j-f^m_j(\beta,T,z))  \nn \\
&+& \sum_k (f_k-f^m_k(\beta,T,z))^2/\sigma_k^2
\ea
where the subscripts $i$ and $j$ span the three BLAST bands and the
subscript $k$ spans the 70 and 160 $\mu$m {\sl Spitzer} bands.  In this
equation, $\sigma^{-1}_{ij}$ is the inverse of the BLAST data
covariance matrix, $\sigma_{k}$ is the error on the $k$th {\sl
Spitzer} flux, $f_i$ is the flux measured in the $i$th band and
$f^m_i$ is the corresponding model flux computed by integrating the
trial SED over the $i$th band response function. The total
far-IR/submm luminosities quoted in the results that follow were
computed by integrating the best fit SED over the wavelength range 10
- 2000 $\mu$m.

We have ignored the effects of flux boosting on the properties derived
from the SED fits. In reality, since the BLAST differential source
count density falls very rapidly with flux \citep[approximately
following ${\rm d}N/{\rm d}S\propto S^{-3.5}$ -- see][]{patanchon09},
the effects of Eddington bias and source confusion mean that the
fluxes in our source catalog are slightly boosted. However, by
restricting our analysis to a relatively bright subset of the BLAST
source catalog and also extracting flux densities at the radio or MIPS
source positions, we expect any boosting to be manageable. In fact,
repeating our analysis on a preliminary de-boosted dataset indicates a
small flux bias which results in a small scatter with negligible bias
to the derived far-IR luminosities and a slight bias of approximately
$+2$K to the dust temperatures.  A full treatment of flux boosting in
the BLAST data will be explored in future work.

\section{Results}
\label{sec_results}


Several of the results presented in this section, where stated, are
limited to a robust sample of sources. We have defined this robust
sample based on the results of the SED fitting discussed below. The
robust sample is therefore a subset of those sources with photometric
redshifts. 

A source is considered robust if its best fit rest-frame dust
temperature is not at the extremes of our uniform temperature prior
(i.e., if $10{\rm K} < T < 50{\rm K}$) and if its SED can be fit with
$\chi^2 \leq N_{dof}+2.71$, where $N_{dof}$ is the number of degrees
of freedom of the fit.  For normally distributed errors, this
threshold in $\chi^2$ corresponds to excluding the worst fit 10\% of
SEDs.  For almost all of our sources, $N_{dof}=3$ since we have five
flux measurements and two SED parameters (normalization and $T$). With
these criteria, our robust sample contains 74 BLAST sources.

Defining the robust sample in terms of the SED fitting is motivated by
the fact that source redshifts are fixed at the value of the
photometric redshift estimated for the counterpart.  If the assumed
redshift is incorrect, either because the photometric redshift is
intrinsically unreliable or because the BLAST source has been
spuriously identified with the counterpart, then this will manifest
itself either with a poor SED fit or by requiring an extreme
temperature to obtain an acceptable fit. 

\subsection{Identifications}
\label{sec_IDs}

Table \ref{tab_src_props} lists the BLAST sources within the FIDEL
region with identified radio and/or 24 $\mu$m counterparts. Table
\ref{tab_src_props_outer} lists the radio and 24 $\mu$m counterparts to
the $\geq 5 \sigma$ BLAST sources located outside the FIDEL region.

Within the FIDEL region the 24 $\mu$m identification rate ($P\leq0.05$)
is 23/78 compared with the radio identification rate of the VLA data
of 29/78. The overall identification rate, i.e., a BLAST source being
identified with $P\leq0.05$ by either a radio or 24 $\mu$m source is
39/78 in the FIDEL area. Splitting this by band, the overall rates (as
a fraction of sources detected at $\geq 5 \sigma$ at each wavelength)
are 31/42 at 250 $\mu$m, 26/45 at 350 $\mu$m and 16/48 at 500 $\mu$m. In
the area outside the FIDEL region, the 24 $\mu$m identification rate of
the SWIRE data is 131/268 compared with the rate of the Norris et
al. data (and the outer part of the VLA data) of 74/220.

Figure \ref{id_rate} plots the overall identification rate as a
function of 250 and 500 $\mu$m flux for sources detected at $\geq
5\sigma$ at each wavelength. The figure shows a small decline in the
identification rate of BLAST sources towards fainter 250 $\mu$m fluxes
and a stronger decline towards fainter 500 $\mu$m fluxes. Over the
full area common to the radio and 24 $\mu$m data, this decline
corresponds to 37 out of 150 of the 250 $\mu$m sources not
being identified and 94 out of 146 of the 500 $\mu$m sources
not being identified. Despite benefitting from a negative K-correction
(which is quite weak at 250 $\mu$m), lower flux BLAST sources are more
likely to reside at higher redshifts. The measured decrease in
identification rate at lower fluxes therefore implies firstly that the
unidentified sources lie at higher redshifts than the identified
sources on average and secondly that there is a larger fraction of
unidentified high redshift 500 $\mu$m sources. Section
\ref{sec_z_models} discusses this further.

By summing the values of $P$ determined in the identification process,
the number of expected false counterparts in the sample can be
estimated. Out of all 103 radio counterparts identified with
$P_r<0.05$, $\sim 1$ of these is expected to be false. Similarly, out
of all 154 24 $\mu$m sources with $P_{24}<0.05$, $\sim 2$ counterparts
would be expected to be false. Limiting this to our robust sample of
counterparts, the expected number of false identifications is $\sim 1$
out of 46 in the radio and $\sim 1$ out of 56 at 24 $\mu$m. 

Relaxing the threshold on $P$ to 0.1 in the area outside the FIDEL
region results in an additional 29 counterparts. Summing $P$ over
these 29 counterparts indicates that two of them would be expected
to be chance alignments. On this basis, we have included counterparts
up to $P\leq 0.1$ in Table \ref{tab_src_props_outer}. However, these
additional sources are only listed in the table and are not
included in any analysis in this paper.

\begin{deluxetable*}{cccccccccccccccc}
\tabletypesize{\scriptsize}
\tablecaption{Radio and 24 $\mu$m counterparts to the $\geq 5 \sigma$ BLAST
sources in the FIDEL region
\label{tab_src_props} }
\tablecomments{
(see Table \ref{tab_src_props_outer} for
the counterparts outside FIDEL). Reading from the left, columns are:
radio co-ordinates, radio flux, $f_r$/mJy, probability of radio
counterpart being a chance alignment, $P_r$, radio radial offset,
$d_{\rm r}/$arcsec, 24 $\mu$m co-ordinates, 24 $\mu$m flux,
$f_{24}$/mJy, probability of 24 $\mu$m counterpart being a chance
alignment, $P_{24}$, 24 $\mu$m radial offset, $d_{24}/$arcsec,
photometric redshift from Rowan-Robinson et al., $z_{\rm RR}$,
COMBO-17 photometric redshift, $z_{\rm 17}$, best fit temperature,
$T$/K, total far-IR/submm luminosity, $L_{\rm FIR}/(10^{10}L_\odot)$
and $\chi^2$ of the SED fit.}
\tablehead{
  \colhead{ID} &
  \colhead{$\alpha$\,(radio)} &
  \colhead{$\delta$\,(radio)} &
  \colhead{$f_{\rm r}$} &
  \colhead{$P_{\rm r}$} &
  \colhead{$d_{\rm r}$} &
  \colhead{$\alpha$\,(24 $\mu$m)} &
  \colhead{$\delta$\,(24 $\mu$m)} &
  \colhead{$f_{24}$} &
  \colhead{$P_{24}$} &
  \colhead{$d_{24}$} &
  \colhead{z$_{\rm RR}$} &
  \colhead{z$_{\rm 17}$} &
  \colhead{$T$} &
  \colhead{$L_{\rm FIR}$} &
  \colhead{$\chi^2$} \\
}
\startdata
    4 &   53.14622 &  -27.92569 &  0.06 &     0.0082 &   1.9 &   53.14611 &  -27.92573 &  5.150 &     0.0006 &   2.0 &  0.06 &  0.04 &  23.3$^{+  1.2}_{-  1.2}$ &  1.47 &  13.6 \\
    6 &   53.12452 &  -27.74028 &  0.34 &     0.0242 &  10.4 &   53.12444 &  -27.74005 &  6.930 &     0.0022 &   9.5 &  0.05 &  0.07 &  25.8$^{+  0.8}_{-  0.8}$ &  2.58 &  1.99 \\
    7 &   53.20818 &  -27.57581 &  0.31 &     0.0361 &  12.7 &      -     &      -     &   -    &     -      &   -   &   -   &  0.23 &  26.2$^{+  1.6}_{-  1.2}$ &  20.8 &  12.9 \\
   17 &   53.20553 &  -27.97914 &  0.21 &     0.0177 &   6.5 &   53.20553 &  -27.97916 &  0.333 &     0.1036 &   6.6 &   -   &  1.23 &  34.6$^{+  2.4}_{-  2.0}$ &  399. &  4.79 \\
   22 &   52.96724 &  -27.65742 &  0.99 &     0.0158 &  11.0 &   52.96697 &  -27.65738 &  0.424 &     0.1808 &  11.9 &   -   &  0.39 &  16.1$^{+  1.6}_{-  1.2}$ &  16.0 &  3.68 \\
   23 &   53.24674 &  -27.72369 &  0.16 &     0.0479 &  11.1 &   53.24668 &  -27.72363 &  0.324 &     0.2288 &  10.9 &   -   &  1.12 &  29.8$^{+  3.2}_{-  2.8}$ &  241. &  1.70 \\
   24 &   52.87452 &  -27.95632 &  0.15 &     0.0196 &   5.9 &   52.87456 &  -27.95623 &  0.275 &     0.1184 &   5.9 &   -   &   -   &             -             &   -   &   -   \\
   26 &   53.19183 &  -27.96262 &  0.07 &     0.0083 &   2.1 &   53.19150 &  -27.96259 &  2.420 &     0.0012 &   2.0 &   -   &  0.11 &  23.3$^{+  1.6}_{-  1.6}$ &  2.47 &  0.56 \\
   34 &   52.95710 &  -27.72408 &  0.15 &     0.0332 &   8.5 &   52.95720 &  -27.72392 &  1.040 &     0.0357 &   9.0 &  0.47 &  0.56 &  22.1$^{+  2.0}_{-  2.0}$ &  41.3 &  2.16 \\
   35 &   53.07106 &  -27.97958 &  3.04 &     0.0228 &  19.0 &   53.07101 &  -27.97958 &  0.064 &     0.8407 &  19.0 &   -   &  1.17 &  32.2$^{+  3.6}_{-  3.2}$ &  256. &  3.93 \\
   36 &   53.32402 &  -27.76846 &  0.11 &     0.0081 &   2.9 &   53.32407 &  -27.76831 &  0.471 &     0.0152 &   2.9 &   -   &  1.56 &  29.4$^{+  2.4}_{-  2.0}$ &  461. &  1.39 \\
   42 &   52.93913 &  -27.77780 &  0.06 &     0.0387 &   5.0 &   52.93900 &  -27.77783 &  0.308 &     0.0811 &   5.3 &   -   &  0.95 &  33.0$^{+  3.6}_{-  3.6}$ &  127. &  11.1 \\
   43 &   53.29048 &  -27.80045 &  0.34 &     0.0628 &  19.7 &   53.29046 &  -27.80044 &  2.590 &     0.0497 &  19.6 &   -   &  0.16 &  28.2$^{+  1.6}_{-  2.0}$ &  5.82 &  13.2 \\
   54 &      -     &      -     &   -   &     -      &   -   &   52.96448 &  -27.74109 &  0.081 &     0.0414 &   1.4 &   -   &  0.94 &  24.1$^{+  2.4}_{-  2.4}$ &  127. &  1.15 \\
   55 &   52.87458 &  -27.93354 &  0.14 &     0.0094 &   3.7 &      -     &      -     &   -    &     -      &   -   &   -   &  1.83 &  47.6$^{+  2.4}_{-  5.7}$ &  952. &  20.1 \\
   63 &   53.31882 &  -27.84430 &  0.11 &     0.0256 &   6.0 &   53.31881 &  -27.84427 &  3.590 &     0.0037 &   6.1 &   -   &  0.11 &  27.0$^{+  1.2}_{-  1.2}$ &  2.51 &  7.00 \\
   66 &   53.02044 &  -27.77983 &  0.09 &     0.0138 &   3.5 &   53.02031 &  -27.77980 &  0.591 &     0.0186 &   3.9 &   -   &  1.16 &  30.2$^{+  3.2}_{-  2.8}$ &  229. &  3.89 \\
   68 &   52.94418 &  -27.95975 &  0.41 &     0.0054 &   4.4 &   52.94399 &  -27.95965 &  1.250 &     0.0056 &   3.7 &  0.26 &  0.36 &  32.6$^{+  2.0}_{-  1.6}$ &  40.9 &  1.74 \\
   85 &      -     &      -     &   -   &     -      &   -   &   52.97289 &  -27.83057 &  0.032 &     0.0119 &   0.5 &  1.07 &  0.73 &  19.3$^{+  2.0}_{-  2.0}$ &  59.0 &  2.81 \\
  102 &   52.85398 &  -27.86892 & 12.79 &     0.0040 &  10.7 &   52.85364 &  -27.86877 &  0.381 &     0.1525 &   9.6 &  1.28 &  1.18 &  41.9$^{+  4.4}_{-  4.4}$ &  357. &  2.80 \\
  110 &   53.07441 &  -27.84972 &  0.05 &     0.0725 &   7.6 &   53.07444 &  -27.84973 &  1.140 &     0.0229 &   7.7 &  0.43 &  0.12 &  23.3$^{+  2.4}_{-  2.4}$ &  1.90 &  0.68 \\
  112 &   53.17499 &  -27.63874 & 29.46 & $<10^{-4}$ &   2.3 &      -     &      -     &   -    &     -      &   -   &   -   &  0.83 &  31.8$^{+  6.1}_{-  4.4}$ &  98.9 &  5.68 \\
  125 &   53.12247 &  -27.58556 &  0.08 &     0.0338 &   5.7 &      -     &      -     &   -    &     -      &   -   &  1.14 &  0.43 &  41.1$^{+  4.4}_{-  4.8}$ &  259. &  10.5 \\
  131 &   53.00352 &  -27.59926 &  2.03 &     0.0139 &  13.4 &      -     &      -     &   -    &     -      &   -   &  0.99 &  0.92 &  25.4$^{+  4.8}_{-  4.4}$ &  102. &  10.0 \\
  132 &   53.10444 &  -27.63972 &  0.11 &     0.0374 &   7.9 &   53.10435 &  -27.63956 &  0.432 &     0.0943 &   7.8 &   -   &  0.62 &  22.5$^{+  3.2}_{-  2.8}$ &  38.9 &  0.78 \\
  136 &   53.11902 &  -27.59375 &  0.09 &     0.0717 &  10.4 &   53.11897 &  -27.59353 &  1.080 &     0.0462 &  11.2 &  0.56 &  0.69 &  28.6$^{+  3.6}_{-  3.6}$ &  74.6 &  0.20 \\
  145 &      -     &      -     &   -   &     -      &   -   &   53.04655 &  -27.98295 &  1.380 &     0.0081 &   5.0 &  0.25 &  0.23 &  20.9$^{+  2.8}_{-  2.4}$ &  7.00 &  3.25 \\
  162 &   52.97921 &  -27.73632 &  0.28 &     0.0356 &  11.8 &   52.97908 &  -27.73624 &  0.131 &     0.5468 &  11.3 &  0.77 &  1.22 &  32.6$^{+  4.4}_{-  4.0}$ &  207. &  0.20 \\
  167 &   53.19950 &  -27.70914 &  0.19 &     0.0427 &  11.4 &   53.19949 &  -27.70913 &  1.070 &     0.0486 &  11.3 &   -   &  0.98 &  21.3$^{+  2.8}_{-  2.4}$ &  84.5 &  2.19 \\
  174 &      -     &      -     &   -   &     -      &   -   &   53.12323 &  -27.66337 &  0.021 &     0.0248 &   0.6 &   -   &  0.95 &  32.2$^{+  6.9}_{-  6.1}$ &  65.0 &  15.1 \\
  179 &   53.24705 &  -27.59286 &  0.19 &     0.0154 &   5.6 &   53.24717 &  -27.59282 &  0.557 &     0.0338 &   5.3 &   -   &  1.12 &  30.6$^{+  5.3}_{-  4.0}$ &  111. &  8.22 \\
  198 &   53.06759 &  -27.65860 &  0.08 &     0.0913 &  12.4 &   53.06747 &  -27.65842 &  1.490 &     0.0326 &  12.0 &  0.37 &  1.34 &  22.1$^{+  4.4}_{-  3.6}$ &  11.9 &  3.17 \\
  218 &   52.92426 &  -27.92695 &  0.10 &     0.0323 &   6.7 &   52.92403 &  -27.92716 &  0.191 &     0.2590 &   7.5 &   -   &  0.91 &  25.4$^{+  3.2}_{-  2.8}$ &  86.0 &  3.07 \\
  221 &   53.04864 &  -27.62401 &  4.05 &     0.0048 &   9.0 &   53.04847 &  -27.62387 &  1.830 &     0.0181 &   8.7 &   -   &  1.55 &  24.5$^{+  4.0}_{-  3.2}$ &  196. &  4.32 \\
  235 &   53.26131 &  -27.94520 &  0.27 &     0.0367 &  11.8 &   53.26065 &  -27.94578 &  0.204 &     0.4137 &  11.7 &   -   &   -   &             -             &   -   &   -   \\
  240 &      -     &      -     &   -   &     -      &   -   &   53.27565 &  -27.73757 &  0.327 &     0.0019 &   0.6 &   -   &  0.84 &  30.6$^{+  4.8}_{-  4.0}$ &  102. &  3.49 \\
  250 &   52.91480 &  -27.68874 &  0.20 &     0.0319 &   9.4 &   52.91470 &  -27.68874 &  1.460 &     0.0215 &   9.1 &  0.94 &  2.09 &  24.9$^{+  4.0}_{-  3.6}$ &  75.3 &  1.73 \\
  262 &   53.18002 &  -27.92068 &  0.07 &     0.0373 &   5.9 &   53.17995 &  -27.92065 &  0.539 &     0.0388 &   5.7 &   -   &  1.32 &  26.6$^{+  3.2}_{-  3.2}$ &  150. &  3.37 \\
  265 &      -     &      -     &   -   &     -      &   -   &   52.86584 &  -27.74164 &  0.048 &     0.0096 &   0.5 &   -   &   -   &             -             &   -   &   -   \\
\enddata
\end{deluxetable*}

\subsection{Redshifts}
\label{sec_redshifts}

\begin{figure}
\epsfxsize=80mm
{\hfill
\epsfbox{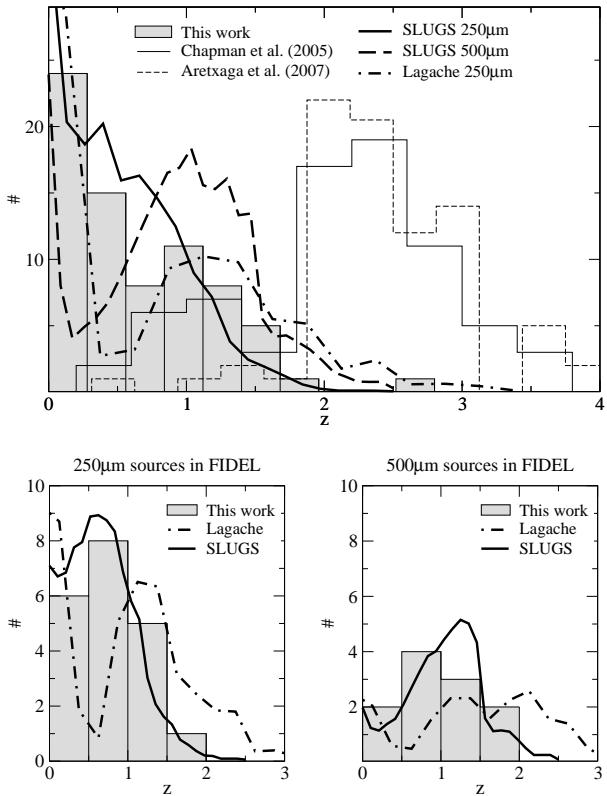}
\hfill}
\caption{{\em Top panel}: The shaded histogram shows the distribution
of photometric redshifts for the 74 BLAST sources in our robust
sample.  Where a counterpart has a redshift estimate from both
COMBO-17 and the catalog of \citet{mrr08}, the redshift that gives the
best modified black body SED fit to the BLAST and {\sl Spitzer} flux is
taken. The two open histograms with the thin continuous and dashed
lines respectively show the distribution of submm sources detected at
850 $\mu$m by \citet{chapman05} and \citet{aretxaga07}. The thick
continuous and thick dashed lines show the predicted redshift
distribution for 250 $\mu$m sources and 500 $\mu$m sources respectively,
computed by evolving the SLUGS local 850 $\mu$m luminosity function of
\citet{dunne00} using the model of \citet{mrr01}.  {\em Bottom left panel}:
The redshift distribution of robust sources detected at 250 $\mu$m in
the FIDEL region (shaded histogram). Over-plotted is the distribution
of 250 $\mu$m sources predicted by our evolved SLUGS model (continuous
line) and by the models of \citet[][dot-dashed line]{lagache04}.  {\em
Bottom right panel}: The same as the bottom left panel but for robust
500 $\mu$m sources in the FIDEL region.
\label{z_dist_comb}}
\end{figure}

Figure \ref{z_dist_comb} shows the distribution of redshifts for our
robust sample of 74 BLAST sources. The figure also shows the redshift
distribution of the SCUBA sources detected at 850 $\mu$m by
\citet{chapman05} and \citet{aretxaga07}.  The submm sources in our
robust sample are clearly located at significantly lower redshifts
than those detected by SCUBA.  The median redshift of the sample of 73
SCUBA sources of \citet{chapman05} lies at $z=2.2$ with an
inter-quartile range of $1.7-2.8$ and the sample of 120 SCUBA sources
of \citet{aretxaga07} has a median of $z=2.4$ and inter-quartile range
of $1.8-3.1$.  In comparison, the distribution of redshifts of our
robust sample has a median of $\sim 0.6$ and an inter-quartile range
of $0.2 - 1.0$.

\begin{figure}
\epsfxsize=75mm
{\hfill
\epsfbox{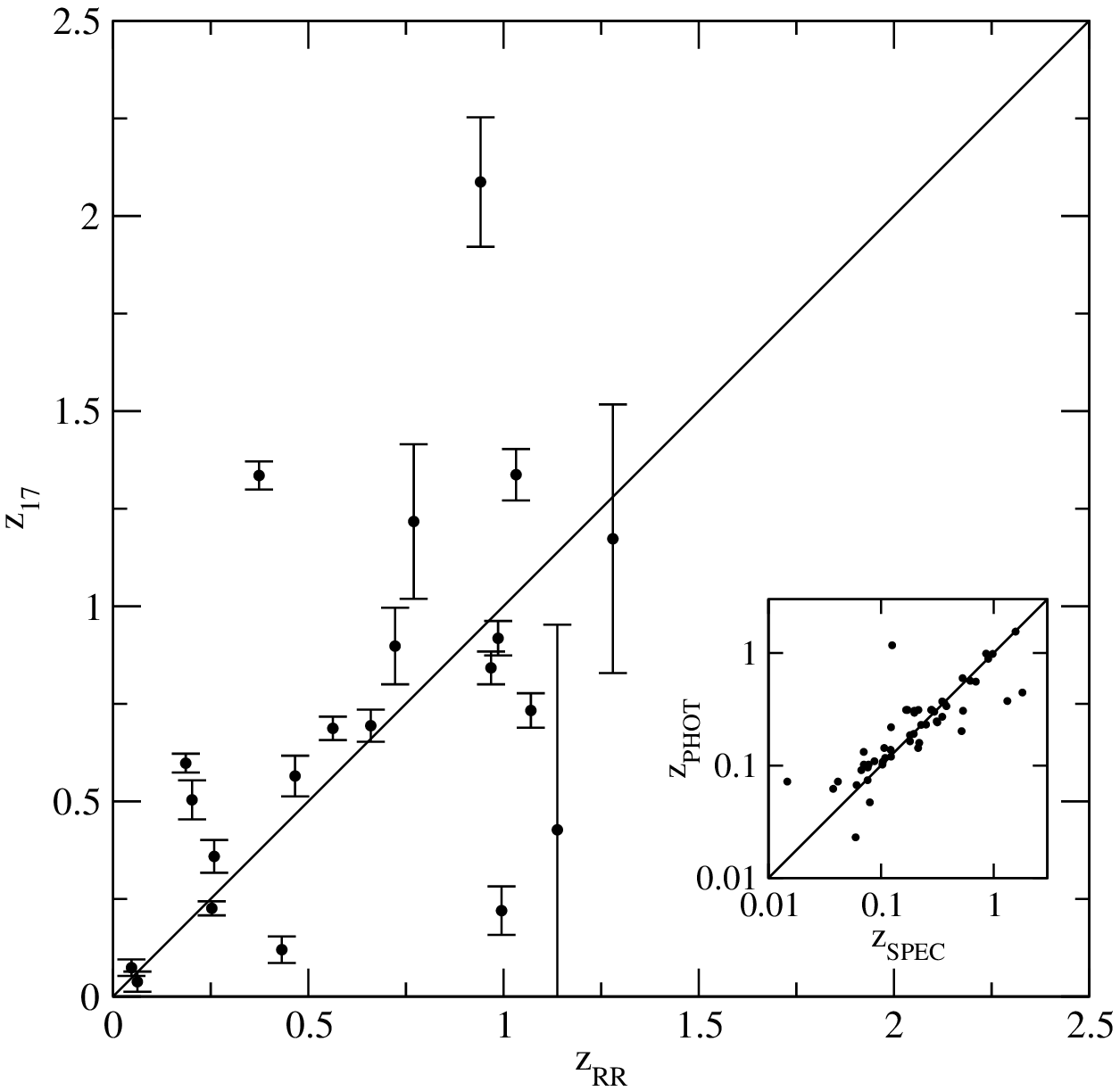}
\hfill}
\caption{{\em Main panel}: Comparison of \citet{mrr08} and COMBO-17 
photometric redshifts for 21 of the $\geq 5\sigma$ BLAST sources where
both redshifts are available. {\em Inset panel}: Comparison of our
recently obtained spectroscopic redshifts \citep{eales09}
with the 53 photometric redshifts in common with the
current work. Where two photometric redshifts exist for a given
source, we have taken that which gives the best SED fit to the
observed submm/far-IR data. Only three out of the 53 photometric
redshifts are discrepant with the spectroscopic redshifts
at the $> 2\sigma$ level.
\label{z_cf}}
\end{figure}

Figure \ref{z_cf} shows a comparison of the optical photometric
redshifts for all sources that have a redshift provided by both
\citet{mrr08} and COMBO-17. The agreement is mediocre with a third of
the 21 redshifts inconsistent at the $> 3 \sigma$ level (although
without uncertainties on the Rowan-Robinson redshifts the number of
inconsistencies quoted here is an upper limit). However, in fitting
the SEDs, for cases where both a Rowan-Robinson and COMBO-17 redshift
exists, we took the redshift which gives the best fit to the
submm/far-IR data. Furthermore, by limiting our analysis of the
derived source properties to the robust sample, the majority of
inaccurate photometric redshifts will be ruled out.

We also very recently obtained spectroscopic redshifts for
approximately half of our radio and/or 24 $\mu$m identified $\geq
5\sigma$ BLAST sources \citep{eales09}. Preliminary analysis has
indicated excellent agreement with the photometric redshifts we have
used in the current work. The inset panel in Figure
\ref{z_cf} shows this comparison for the 53 sources in common and
indicates that our method of selecting the photometric redshift that
best fits the observed fluxes when two redshifts are available works
well. Although we have not incorporated the spectroscopic redshifts in
the analysis carried out in this paper, this excellent agreement
increases the confidence in the source properties that follow from the
SED fitting, since these strongly depend on the redshift assumed.

\subsubsection{Comparison with model predictions}
\label{sec_z_models}

We have predicted the redshift distribution of the BLAST sources
using an empirical model \citep{eales09} derived from the
results of the SCUBA Local Universe and Galaxy Survey 
\citep[SLUGS;][]{dunne00,vlahakis05}. The model is based on the
sample of 104 galaxies observed by \citet{dunne00} using SCUBA
at 450 and 850 $\mu$m. The galaxies form a
statistically-complete sample above a $60\mu$m flux limit of 
5.24 Jy and constitute the only large sample of galaxies with
empirical spectral energy distributions that stretch from the
far-IR to submm wavebands, which is a major advantage
over existing models for submm surveys.

We can use the SLUGS sample to predict the source counts at any
frequency in a straightforward way.  The number of sources above a
given 250 $\mu$m flux density is given by
\be
\label{eq_no_counts_integral}
N(>f_{250}) = \sum_{j=1}^{104} 
\int_0^{z(L_j(z),f_{250})} {1 \over V_j} \, dV
\ee
where $V_j$ is the comoving volume in which the $j$th SLUGS source
could have been detected in the original sample from which it was
selected, and the integral is over comoving volume out to the redshift
at which the source would be detected in the current sample. We assumed
that the luminosity of the $j$th SLUGS galaxy is given by
\be
L_j(z) = E(z) L_j(0)
\ee
where $L_j(0)$ is the empirical luminosity of the $j$th SLUGS galaxy
at the appropriate rest-frame wavelength and $E(z)$ is an evolution
function. We have implicitly assumed `luminosity evolution' rather than
`number-density evolution', which is necessary anyway because it is
impossible to fit both the cosmic background radiation and the
submm source counts with number-density evolution.  In
practice, we have used the simple luminosity-evolution model from
\citet{mrr01}, in which luminosity is given by
\be
L(t) = L(t_0) \left( {t \over t_0} \right)^D \, 
\exp\left[C \, (1-t/t_0)\right] 
\ee
where $t$ is the time from the big bang and $t_0$ is the time at the
current epoch. $C$ and $D$ are parameters of the model. We have found 
that values of $C=9$ and $D=3$ give acceptable fits to the spectral 
shape and intensity of the cosmic background radiation and to the SCUBA 
850 $\mu$m and {\sl Spitzer} 70 $\mu$m source counts. By adjusting the 
integral limits in equation (\ref{eq_no_counts_integral}), the
distribution of redshifts can be computed.

The predicted redshift distribution for the BLAST galaxies detected at
$\geq 5\sigma$ at 250 $\mu$m and at 500 $\mu$m using this approach is
shown in the top panel of Figure \ref{z_dist_comb}. In addition, the
figure shows the redshift distribution according to the galaxy
evolution model of \citet{lagache04}.  The predictions apply
specifically to the BLAST CDFS survey, taking into account the
different areas and depths of BGS-Wide and BGS-Deep.

To allow for possible biases in our BLAST catalogue caused by flux
boosting and source confusion, we have disregarded absolute model
normalizations throughout this section and have instead normalized the
model counts to match the BLAST counts. In this way, our aim is to
compare only distribution morphologies. We refer the reader to
\citet{patanchon09} for an account of the BLAST number counts and the
subsequent implications for models.

A more direct comparison of the model redshift distributions is shown
in the bottom two panels of Figure \ref{z_dist_comb}. Here, we have
limited the comparison to the FIDEL area where the BLAST, radio and 24
$\mu$m data are deepest and most uniform. The bottom left panel shows
the redshift distribution of the 20 robust sources detected at 250
$\mu$m. Similarly, the bottom right panel shows the redshift
distribution of the 11 robust sources detected at 500 $\mu$m.  Both
lower panels show the corresponding distributions predicted by our
evolved SLUGS model and the Lagache et al. model.

\begin{figure}
\epsfxsize=80mm
{\hfill
\epsfbox{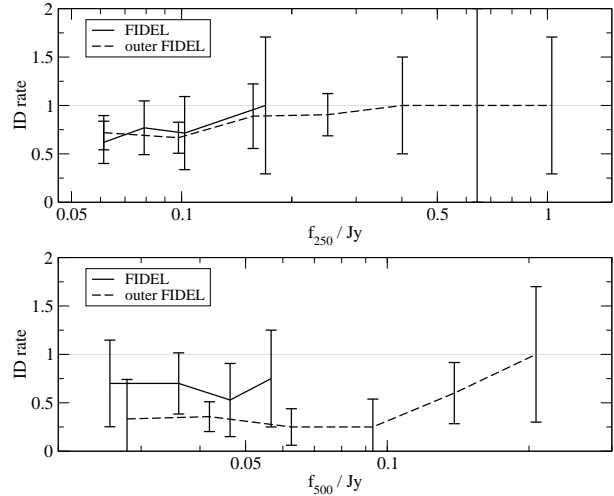}
\hfill}
\caption{The identification rate of BLAST sources (having
either a radio or 24 $\mu$m counterpart) as a function of 250 $\mu$m
flux (top) and 500 $\mu$m flux (bottom). In both plots, the continuous
line corresponds to the FIDEL region and the dashed line to the area
outside FIDEL common to the radio and 24 $\mu$m data. The plots
include only those sources detected with significance $\geq 5\sigma$ 
at each wavelength.
\label{id_rate}}
\end{figure}

The Lagache et al. and evolved SLUGS model distributions show
considerable differences. In the 250 $\mu$m case, the largest
discrepancy occurs at $z<1$. The Lagache et al. model predicts a
deficit around $z \simeq 0.5$ at the division between the quiescent
population at $z<0.5$ and the starbursting population at
$z>0.5$. However, according to the evolved SLUGS model, a peak is
anticipated at $z \simeq 0.5$.  The measured distribution more closely
resembles the SLUGS model although the large uncertainty due to
Poisson noise means that the data can not be used to unambiguously
validate one model over the other.  In the case of the 500 $\mu$m
sources, there are similar discrepancies between the models. Once
again, the evolved SLUGS model appears more consistent with the
observed redshifts but the dominant Poisson noise precludes
verification of either model.

Despite their differences, neither of the model redshift distributions
in Figure \ref{z_dist_comb} implies that we have failed to identify
counterparts or obtain redshifts for a significant number of $\geq
5\sigma$ sources beyond $z \simeq 2$ where $>80\%$ of the SCUBA
population resides. Of the two models considered, the Lagache et
al. model is most discrepant with our sample, implying that we may
have missed a small fraction of higher redshift sources. This is
indeed consistent with our expectations based on the declining
identification rate of 250 $\mu$m sources toward fainter fluxes (see
Figure \ref{id_rate}). However, neither the Lagache et al. model nor
the SCUBA model predicts a sufficient fraction of high redshift 500
$\mu$m sources required to explain the stronger decline in
identification rate seen at fainter 500 $\mu$m fluxes.

\begin{figure}
\epsfxsize=80mm
{\hfill
\epsfbox{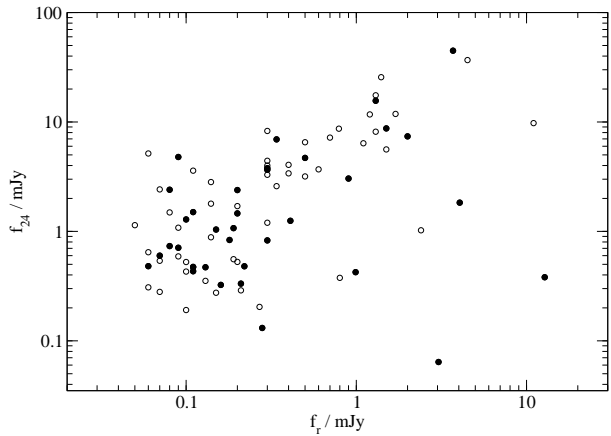}
\hfill}
\caption{Correlation between 1.4GHz radio flux and 24 $\mu$m 
{\sl Spitzer} flux for all counterparts to the $\geq 5 \sigma$ BLAST
sources identified at both wavelengths.  The filled symbols denote
sources in the robust sample (all of which have photometric redshifts).
\label{f24_vs_f_radio}}
\end{figure}

A obvious question that arises is to what degree does the fraction of
sources without redshifts reconcile the differences between the SCUBA
redshift distributions mentioned previously and the redshift
distribution of our robust sample? Would we expect some of these
sources to coincide with the SCUBA population?  Unfortunately, we can
only provide an upper limit. Figure \ref{f24_vs_f_radio} plots the
radio flux versus the 24 $\mu$m flux for all counterparts identified
at both wavelengths.  Sources in the robust sample (which all have
photometric redshifts) are indicated by filled points. The
distribution of non-robust sources (mostly without redshifts) is
clearly not concentrated in any specific part of the $f_r-f_{24}$
plane, in particular at faint fluxes where more distant sources would
be expected to lie. (Although not shown, the same is true when every
source with a redshift is over-plotted).  The rate of assignment of
redshifts within the sub-sample of BLAST sources with counterparts
therefore does not decline at higher redshifts at a detectable
level. This in turn means that if we had been able to obtain redshifts
for all 198 identified counterparts, their redshift distribution would
be similar in shape to that of our robust sample except with greater
normalization. Therefore, even under the extreme assumption that all
153 unidentified sources lie at high redshift ($z\simgreat 2$), there is
still only an overlap of just under 50\% with the SCUBA population.

However, we emphasize that the $\geq 5 \sigma$ sources considered in
this paper are at the bright end of the sources detected by BLAST
within BGS-Deep/Wide.  The fainter sources, generally detected with
lower significance, are expected to probe higher redshifts. This does
indeed seem to be borne out by the detection of a significant fraction
of the extragalactic far-IR/submm background from fainter, higher
redshift BLAST sources \citep{pascale09,marsden09}.

\subsection{Source SEDs}
\label{sec_seds}

Tables \ref{tab_src_props} and \ref{tab_src_props_outer} list the best
fit rest-frame dust temperatures and total far-IR/submm luminosities
for the 114 sources with available photometric redshifts. The SEDs for
those sources located in the FIDEL region are plotted in Appendix
\ref{app_seds}.

In Figure \ref{L_fir_vs_z} we plot the total far-IR/submm luminosity,
$L_{\rm FIR}$, of the BLAST sources in the robust sample against
redshift.  Superimposed on this plot are curves showing the minimum
$L_{\rm FIR}$ a source must have in order to be detected at $5 \sigma$
in one of the three BLAST bands. These thresholds were computed using
the survey sensitivity in BGS-Deep (55, 45 and 30\,mJy/beam at 250,
350 and 500\, $\mu$m respectively to 5$\sigma$) hence sources in the
shallower region will be subjected to a set of curves shifted to
slightly lower redshifts.  The band in which the detection limit
occurs varies as a function of redshift and dust temperature (e.g., a
10 K source meeting the requirement for a detection of $5\sigma$ in
one of the three bands will only ever be met in the 500 $\mu$m
band). The figure shows that for a given $L_{\rm FIR}$ out to a
redshift of $z \simeq 1.6$, cooler sources can be detected at greater
distances.  Beyond this redshift, 10 K sources begin to rapidly move
out of the 500 $\mu$m band and increasingly warmer sources become more
readily detected. However, since the curves for hotter sources (40-50
K) become flatter at higher redshifts, a small increase in temperature
of a hotter source pushes the detection limit to much lower
redshifts. The net result is that there exists a modal mid-range dust
temperature that slowly rises with redshift.

A complication is that the above argument assumes the sources have a
uniform distribution of dust mass. In practice, a cooler source at
high redshift must have substantially more dust than a hotter source
to remain detected. Since sources with more dust are more rare, this
greatly reduces the probability of detecting cooler sources (10-20 K)
out to higher redshifts ($z \simgreat 1.5$). We return to this point
below.

\begin{figure}
\epsfxsize=85mm
{\hfill
\epsfbox{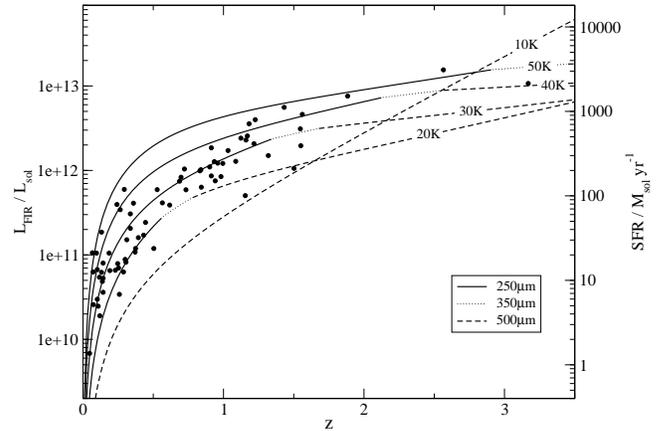}
\hfill}
\caption{Correlation between total far-IR/submm luminosity, $L_{\rm
FIR}$, and redshift for our robust sample of 74 sources. The right
hand ordinate gives the star formation rate (SFR) in M$_\odot$/year
estimated from the scaling: SFR/M$_\odot$\,yr$^{-1}=2.0 \, \times \,
10^{-10} L_{\rm FIR}/L_\odot$ \citep{hughes97}. The different curves
show, for a range of dust temperatures as labelled, the minimum 
$L_{\rm FIR}$ a source must have in order to be detected at
$5 \sigma$ in one of the three BLAST bands. The band in which
this threshold occurs is indicated by the line style displayed in the
legend. The $5 \sigma$ detection limits used are those applicable 
to the BGS-Deep survey region (55, 45 and 30\,mJy/beam at 250, 350 
and 500\, $\mu$m respectively).
\label{L_fir_vs_z}}
\end{figure}

\begin{figure}
\epsfxsize=80mm
{\hfill
\epsfbox{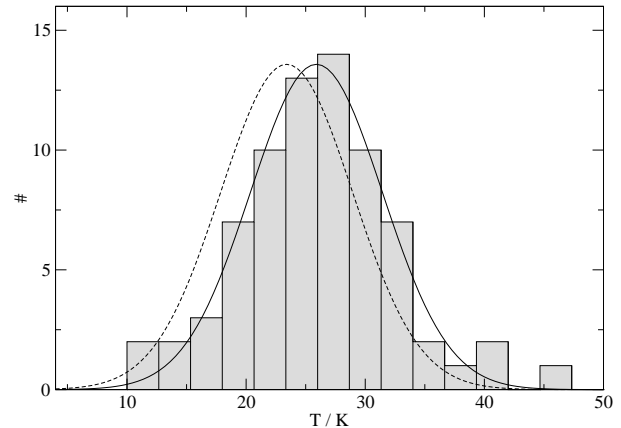}
\hfill}
\caption{Histogram of dust temperatures for our robust sample of 74
sources (see text). The plotted best fit Gaussian (continuous line)
has a $1\sigma$ width of 5.2 K and mean of 25.9 K. Repeating the
analysis but changing $\beta$ from 1.5 to 2.0 gives a very similar
distribution with mean 23.4 K (dashed line).
\label{T_hist}}
\end{figure}

Figure \ref{T_hist} shows the distribution of best fit rest-frame dust
temperatures in our robust sample for the assumed fixed value of
$\beta=1.5$. The distribution is well approximated by a Gaussian of
mean 26 K and $1 \sigma$ width of 5 K.  This spread in temperatures is
comparable to the $1 \sigma$ error on the mean of 4 K determined
through error propagation. Fixing $\beta=2$ instead gives a similar
distribution but shifted to slightly lower temperatures with a mean of
23 K.

Comparing our dust temperatures with those derived for submm sources
detected at 850 $\mu$m by SCUBA, we find that the BLAST sources are
significantly cooler on average. The sample of 73 SCUBA sources of
\citet{chapman05} has a median dust temperature of $36 \pm 6$ K,
\citet{kovacs06} determined a median temperature of $35 \pm 3$ K
for their sample of 15 sources while the sample of 25 SHADES galaxies
studied by \citet{coppin08} was found to have a median temperature of
29 K (but with a large scatter of 18 K). As we discussed in Section
\ref{sec_redshifts}, the SCUBA sources lie at significantly higher
redshifts on average, hence an obvious question is whether our sample
of BLAST sources supports a trend for hotter sources at higher
redshifts.

\begin{figure}
\epsfxsize=80mm
{\hfill
\epsfbox{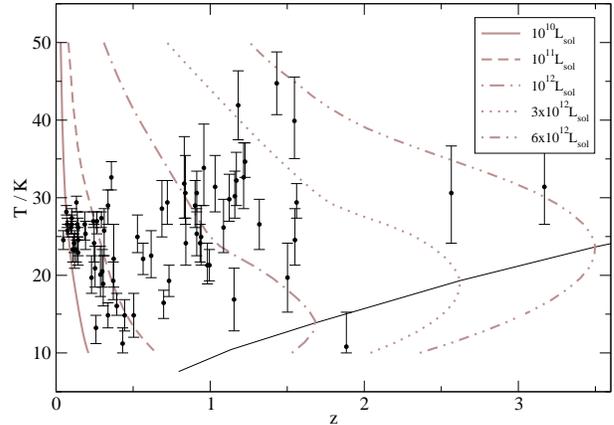}
\hfill}
\caption{Correlation between rest-frame dust temperature, $T$, and
redshift for our robust sample of 74 sources (see text).  The curves
show the sensitivity of the survey to redshift as a function of dust
temperature for a source with a range of total far-IR/submm
luminosities. The curves are determined by requiring a $5 \sigma$
detection of the source in one of the three BLAST bands. The
sensitivity limits used are those of BGS-Deep (55, 45 and 30\,mJy/beam
at 250, 350 and 500\, $\mu$m respectively to 5$\sigma$). The thin
continuous line in the lower part of the plot shows the dust
temperature where the threshold in luminosity for detection is
minimised for a given redshift.  Sources do not lie along this
threshold because it corresponds to high dust masses which are rare.
\label{T_vs_z}}
\end{figure}

Figure \ref{T_vs_z} shows the rest-frame dust temperatures for the
BLAST sources in our robust sample plotted against redshift. The
scatter is large, but it is clear that hotter than average sources lie
at higher redshifts. However, the selection effects discussed
previously must be taken into consideration. The curves in Figure
\ref{T_vs_z} are the sensitivity curves from Figure \ref{L_fir_vs_z}
transformed into the $T-z$ plane.  For a given redshift, the curves in
Figure \ref{T_vs_z} show that there is a dust temperature where the
threshold in luminosity for detection is minimised. This is shown by
the thin continuous line. Since lower luminosity sources are more
common at a given redshift, this minimum threshold luminosity is where
a population of sources with a uniform distribution of temperatures
and dust masses would be expected to lie. Moving to colder or hotter
temperatures away from this threshold at a fixed redshift requires a
higher luminosity for detection. This is one of two dominant selection
effects.

The second dominant selection effect arises from the fact that in
reality, the dust mass distribution is far from uniform.  Moving
toward colder temperatures along a line of constant luminosity in
Figure \ref{T_vs_z} requires the dust mass of a source to increase
rapidly to stay within the survey sensitivity limits. For example,
moving from $T=30$ K to $T=20$ K along the curve corresponding to
$L_{\rm FIR}=10^{12}\,L_\odot$ requires an increase in dust mass from
$\sim 5 \times 10^6 \,{\rm M}_\odot$ to $\sim 5 \times 10^7 \,{\rm
M}_\odot$ (assuming a dust mass opacity coefficient of $0.1 {\rm
m}^2/{\rm kg}$).  Using the dust mass function of local submm galaxies
determined by \citet{vlahakis05}, there are $\sim 10$ times as many
galaxies with a dust mass of $5 \times 10^6 \,{\rm M}_\odot$ than with
a dust mass of $5 \times 10^7 \,{\rm M}_\odot$. This required dust
mass at 20 K corresponds to the knee in the dust mass function, beyond
which galaxies with more dust become vastly more rare.  Repeating the
estimate for 10 K results in an expected reduction in the number of
galaxies by a factor of $\sim 10^6$. Along the higher luminosity
sensitivity curves, the same principles apply but since dust mass is
proportional to luminosity for a fixed temperature the knee of the
dust mass function and therefore the sudden rarity of galaxies applies
at warmer temperatures.

This strong selection effect pushes sources away from the minimum
threshold luminosity plotted in Figure \ref{T_vs_z}.  The observed
trend of increasing rest-frame dust temperature with redshift is
therefore where the strength of both selection effects are
approximately balanced. Both cause an observed increase in 
temperature with redshift, but the luminosity effect limits
the upper detectable temperature and the dust mass effect limits
the lower temperature.

\begin{figure}
\epsfxsize=80mm
{\hfill
\epsfbox{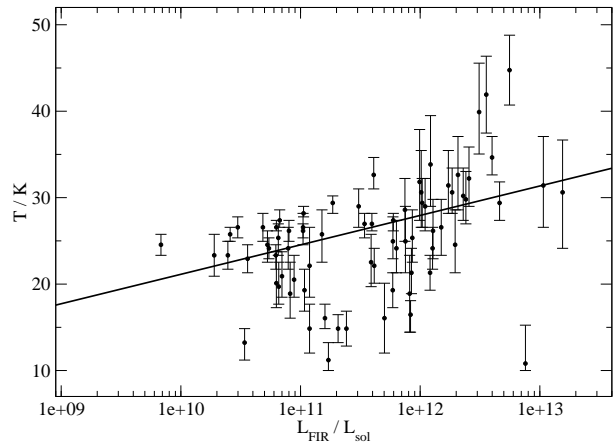}
\hfill}
\caption{Correlation between dust temperature, $T$, and 
total far-IR/submm luminosity, $L_{\rm FIR}$, for our robust
sample of 71 sources. The straight line fit is
$T=-12.9+3.4\,\,{\rm log}_{10} (L_{\rm FIR}/L_\odot)$.
\label{T_vs_L_fir}}
\end{figure}

In Figure \ref{T_vs_L_fir}, we plot the dust temperature of sources in
our robust sample against luminosity.  The straight line fit is
$T=-12.9+3.4\,\,{\rm log}_{10} (L_{\rm FIR}/L_\odot)$.  The increase
in temperature at higher luminosity is a reflection of the trend
discussed previously.  This is consistent with a known positive
correlation between intrinsic dust temperature and far-IR luminosity
for luminous infrared galaxies in the local Universe
\citep{chapman03,chapin09}. A more complete investigation will be
discussed in forthcoming work (Chapin et al., in preparation).

\section{Summary and Discussion}
\label{sec_summary}

We have identified counterparts detected at 1.4GHz and/or 24 $\mu$m to
198 of the 350 $\geq 5\sigma$ BLAST sources contained within the $\sim
9$ deg$^2$ of BGS-Wide and BGS-Deep centered on the CDFS.  In the
region within BGS-Deep covered by the deep 24 $\mu$m FIDEL catalog
where we also have deep VLA data, the identification rate is 23/78 for
the 24 $\mu$m sources and 29/78 for the radio sources.  In the $\sim
8$ deg$^2$ surrounding the FIDEL region, the identification rate for
24 $\mu$m SWIRE sources is 131/268 compared with the rate of 74/220
using the Norris et al. radio catalog.

Of the identified counterparts, 114 have photometric redshifts
previously estimated by COMBO-17 and/or \citet{mrr08}.  Using these
redshifts, we have fitted modified black body SEDs to the BLAST fluxes
measured at wavelengths 250, 350 and 500 $\mu$m and {\sl Spitzer}
fluxes at 70 and 160 $\mu$m.  We have defined a robust sample of 74
sources whose SEDs fit the observed fluxes with $\chi^2 \le
N_{dof}+2.71$ (rejecting the worse fit 10\% assuming normal errors)
and whose resulting best fit dust temperatures are not at the extremes
of 10 K or 50 K of our uniform temperature prior.

The distribution of redshifts of our robust sample has a median of
$\sim 0.6$ and an inter-quartile range of $0.2 - 1.0$.  The dust
temperatures are approximately normally distributed with a median
temperature of $T\simeq 26 \pm 5$ K (for $\beta=1.5$, or $T\simeq 23
\pm 5$ K for $\beta=2$) and the distribution of bolometric far-IR/submm
luminosity has a median of $4\times10^{11} L_\odot$.

Comparing the $\geq 5 \sigma$ submm sources detected by BLAST with
those detected by SCUBA at 850 $\mu$m, we find a stark contrast. It is
clear from the redshift distributions alone that the BLAST sources in
our robust sample are a significantly less distant population.  The
median redshift of submm sources detected to date in 850 $\mu$m SCUBA
surveys lies somewhere between $z=2-2.5$ \citep[e.g.,][]
{chapman05,aretxaga07,younger07}.  In comparison, 75\% of the robust
BLAST sources, for which we have identified counterparts in a similar
manner to the SCUBA surveys \citep[i.e., with a combination of radio
and 24 $\mu$m data, the same used by][for example]{ivison07}, lie at
$z<1$.

This significant difference can not be explained by our selection of
robust sources. Within the subset of BLAST sources with identified
counterparts, the robust sources form a uniform random sampling of the
plane spanned by radio and 24 $\mu$m counterpart flux. In other words,
we have not systematically failed to obtain redshifts for a more
distant subset of identified BLAST sources where the radio and/or 24
$\mu$m flux would be expected to be lower on average. Therefore, we
would not expect the redshift distribution of the full sample of 198
identified BLAST sources to extend to much greater redshifts. However,
there remains the possibility that the remaining 153 BLAST sources
without identified counterparts lie at higher redshifts on average. We
have measured a decrease in the identification rate towards fainter
BLAST fluxes which supports this hypothesis. Around a quarter of the
250 $\mu$m sources detected at $\geq 5 \sigma$ remain unidentified
compared to nearly two thirds of 500 $\mu$m sources.

We have compared our measured redshift distributions with predictions
made by the galaxy evolution model of \citet{lagache04} and by our own
model where we evolved the local 850 $\mu$m luminosity function with
the empirical models of \cite{mrr01}. Considering sources detected at
250 $\mu$m, our own model implies that we have not failed to identify
a significant number of 250 $\mu$m sources at higher redshift within
the survey sensitivity limits. However, the Lagache model predicts
that we may have failed to identify a small fraction ($\sim 10-15\%$)
beyond $z \simgreat 2$ which is compatible with our observed small
decline in the identification rate at faint 250 $\mu$m fluxes.
Conversely, at 500 $\mu$m, neither model implies a large enough
fraction of high redshift sources to accommodate the strong decline in
identification rate toward faint fluxes measured. This decline results
in $\sim 100$ unidentified 500 $\mu$m sources. Even extending the
models to include $\geq 3\sigma$ sources only shows a $\sim 25\%$
increase in the number of sources predicted at $z>2$. 

Considering all of the evidence presented (and bearing in mind that
the models are not absolutely normalized), the most likely explanation
is that the unidentified BLAST sources do indeed lie at higher
redshifts than the identified subset on average, but that they also
include a significant fraction of lower redshift sources. Failure to
identify a low redshift source will occur if the counterpart is too
faint, either to the extent that it is not detected or that it yields
a value of $P$ in the ID procedure that falls outside the threshold.
To conclude our discussion of redshifts, although we cannot fully
quantify the overlap of the redshift distribution of BLAST sources
considered in this paper with that of the SCUBA population, we can
place a strong upper limit. Under the extreme assumption that all 153
unidentified sources lie at higher redshift (say $z>2$) than the 198
identified, the overlap with the SCUBA population is still just under
50\%.

We have also found that the average dust temperature of the BLAST
sources is clearly different to that of sources detected in the SCUBA
surveys.  The BLAST sources have a significantly cooler temperature
distribution compared to the higher redshift 850 $\mu$m population. For
example, \citet{chapman05} measured a median dust temperature of $36
\pm 6$ K. This is consistent with a trend observed in our sample
of BLAST sources such that sources with higher dust temperatures are
seen at higher redshifts. We have shown that this trend is the result
of two strong selection effects.

This paper has addressed a small fraction of the analysis made
possible by the multi-wavelength catalog of BLAST sources published
herein and the identification of their radio and 24 $\mu$m
counterparts.  Further analysis will be conducted in forthcoming BLAST
papers. 

An immediate priority is to increase the number of source redshifts.
We have very recently obtained spectroscopic redshifts with the
multi-fiber spectrometer, AAOmega, on the Anglo-Australian telescope
for $>100$ of the BLAST sources \citep{eales09}.
Preliminary results show very close agreement with the
subset of sources in common that have previously derived photometric
redshifts, although the spectroscopic sample extends the total number
significantly. The multi-wavelength catalog (in combination with the
{\sl Spitzer} photometry at 70 and 160 $\mu$m) allows submm photometric
redshifts to be estimated for those sources without counterparts.
With a larger sample of redshifts, more stringent limits can be placed
on galaxy evolution models and the redshift distributions they
predict.  These redshifts also enable investigation of the evolution
of the submm luminosity function (Chapin et al., in preparation) and
more rigorous investigation of the far-IR/radio correlation (Ivison et
al., in preparation).  Another possibility brought about by the
spectra is measurement of equivalent line widths to provide a direct
estimate of unobscured star formation rates.  Combining spectra with
optical/near-IR morphology will give new insight into the types of
systems that BLAST, and hence Herschel (see below), is sensitive
to. Determination of the dominant processes at play will greatly
assist our understanding of the link between these systems and the
local population of galaxies.

The photometry upon which the submm source properties derived in this
paper have been based, the three BLAST bands at 250, 350 and 500
$\mu$m and the two {\sl Spitzer} channels at 70 and 160 $\mu$m, is
almost identical to what the Herschel Space Observatory is expected to
deliver with SPIRE and PACS. Although the 9 deg$^2$ BLAST survey
analyzed here represents a significant leap forward in terms of areal
coverage of submm surveys, a subsequent leap is imminent with the
anticipated several hundred square degree surveys to be conducted with
Herschel.


\appendix
\section{Submm SEDs}
\label{app_seds}

\begin{figure*}
\epsfxsize=160mm
{\hfill
\epsfbox{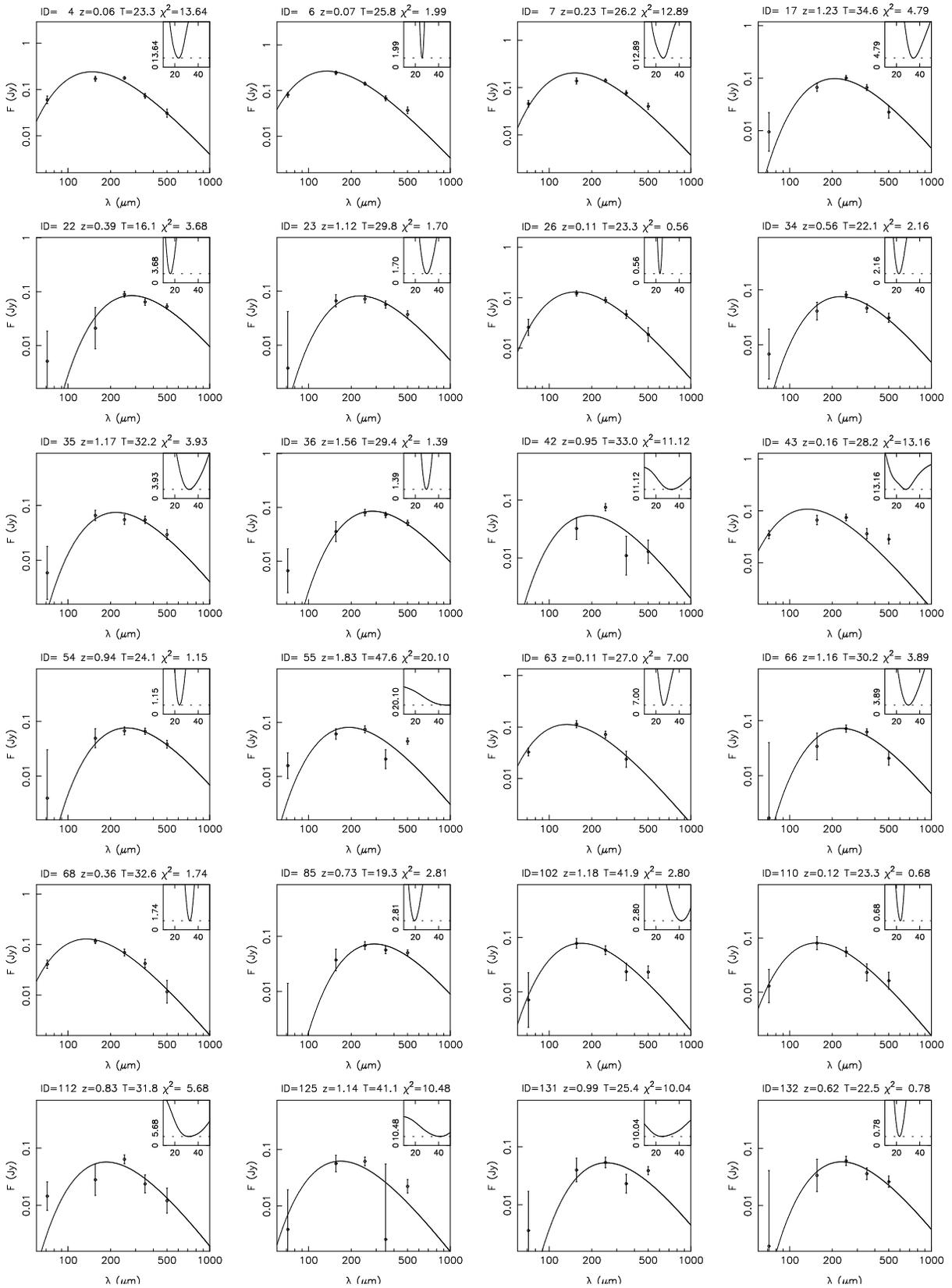}
\hfill}
\caption{Best fit SEDs for $\geq 5\sigma$ BLAST sources in the FIDEL
region. In the minimization, temperature is allowed to vary between
10 K and 50 K with a uniform prior, $\beta$ is fixed at the value 2.0.
Redshifts are fixed at the redshift of the optical counterpart.  Where
more than one redshift exists, that which gives the best fit to the
BLAST data is taken. Inset plot in each panel shows $\chi^2$ versus
$T$.
\label{seds1}}
\end{figure*}

Figure \ref{seds1} shows the best fit SEDs to the BLAST and
{\sl Splitzer} 70 and 160 $\mu$m data for the
$\geq 5\sigma$ BLAST sources in the FIDEL region.

\begin{figure*}
\figurenum{\ref{seds1} {\em continued}}
\epsfxsize=160mm
{\hfill
\epsfbox{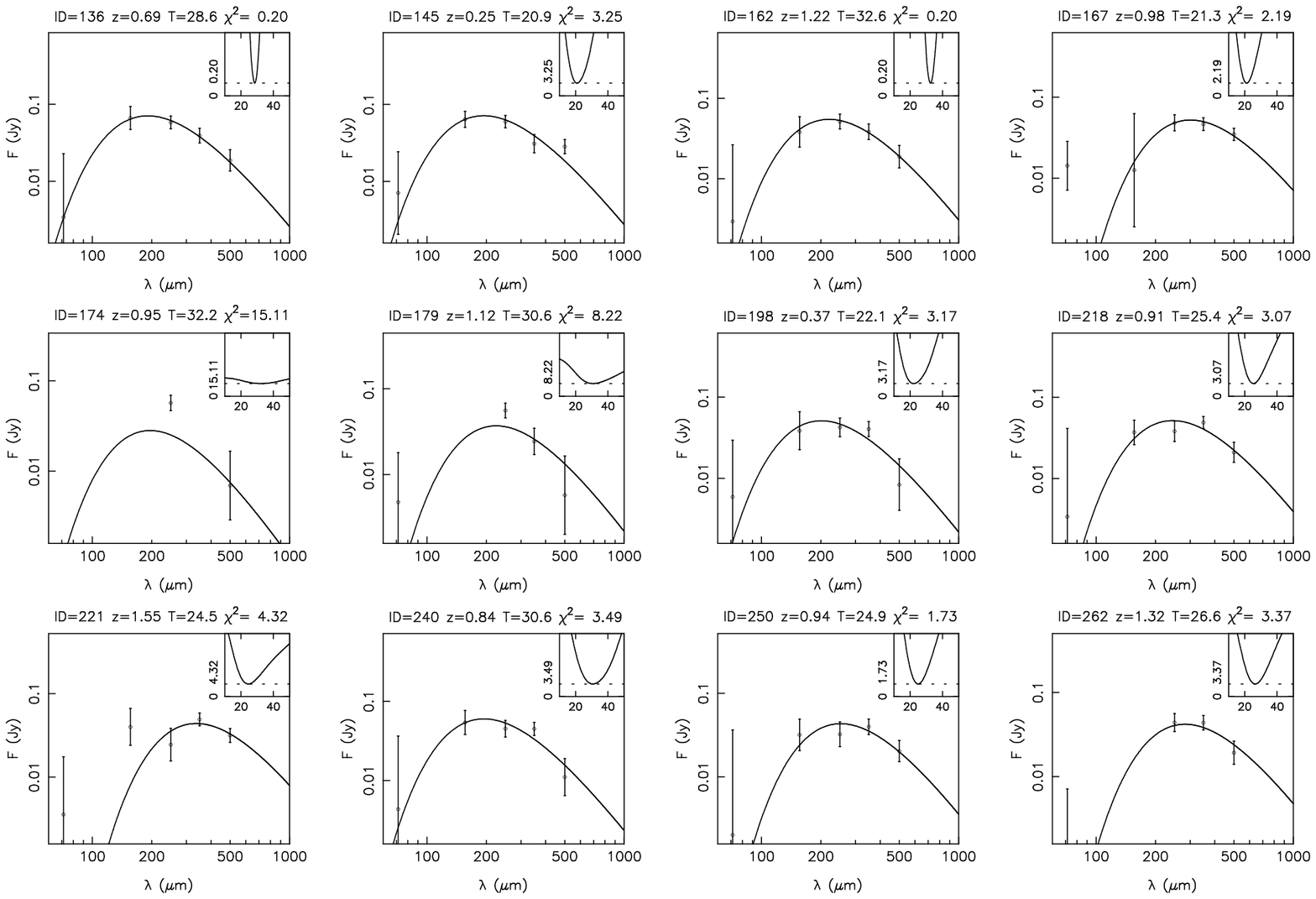}
\hfill}
\caption{Best fit SEDs for $\geq 5\sigma$ BLAST sources in the FIDEL
region.
\label{seds2}}
\end{figure*}

\section{Counterparts outside the FIDEL region}
\label{app_out_fidel}

Table \ref{tab_src_props_outer} lists the radio and/or 24 $\mu$m
counterparts to the BLAST sources outside the FIDEL region.

\begin{deluxetable*}{cccccccccccccccc}
\tabletypesize{\scriptsize}
\tablecaption{
Radio and 24 $\mu$m counterparts to the $\geq 5 \sigma$ BLAST
sources in the BLAST region outside the FIDEL region
\label{tab_src_props_outer} }
\tablecomments{
(see Table \ref{tab_src_props} for counterparts within the FIDEL region). 
Reading from the left, columns are: radio co-ordinates, radio flux,
$f_r$/mJy, probability of radio counterpart being a chance alignment,
$P_r$, radio radial offset, $d_{\rm r}/$arcsec, 24 $\mu$m co-ordinates,
24 $\mu$m flux, $f_{24}$/mJy, probability of 24 $\mu$m counterpart being
a chance alignment, $P_{24}$, 24 $\mu$m radial offset, $d_{24}/$arcsec,
photometric redshift from Rowan-Robinson et al., $z_{\rm RR}$,
COMBO-17 photometric redshift, $z_{\rm 17}$, best fit temperature,
$T$/K, total far-IR/submm luminosity, $L_{\rm FIR}/(10^{10}L_\odot)$
and $\chi^2$ of the SED fit.
Note that unlike Table \ref{tab_src_props}, included here are
counterparts with $0.05 < P \leq 0.1$. This results in an additional 29
counterparts at the expense of an expected extra two false
counterparts. None of these additional counterparts are included
in any of the analysis presented in this paper but are simply
listed here for reference. }
\tablehead{
  \colhead{ID} &
  \colhead{$\alpha$\,(radio)} &
  \colhead{$\delta$\,(radio)} &
  \colhead{$f_{\rm r}$} &
  \colhead{$P_{\rm r}$} &
  \colhead{$d_{\rm r}$} &
  \colhead{$\alpha$\,(24 $\mu$m)} &
  \colhead{$\delta$\,(24 $\mu$m)} &
  \colhead{$f_{24}$} &
  \colhead{$P_{24}$} &
  \colhead{$d_{24}$} &
  \colhead{z$_{\rm RR}$} &
  \colhead{z$_{\rm 17}$} &
  \colhead{$T$} &
  \colhead{$L_{\rm FIR}$} &
  \colhead{$\chi^2$} \\
}
\startdata
    1 &   52.33788 &  -28.13343 &  4.50 &     0.0005 &   3.3 &   52.33796 &  -28.13353 & 36.781 & $<10^{-4}$ &   2.9 &   -   &   -   &             -             &   -   &   -   \\
    2 &   52.48567 &  -28.77572 &  5.40 &     0.0004 &   3.0 &      -     &      -     &   -    &     -      &   -   &   -   &   -   &             -             &   -   &   -   \\
    3 &   51.92112 &  -28.38893 &  3.70 &     0.0015 &   5.8 &   51.92133 &  -28.38897 & 44.827 & $<10^{-4}$ &   5.4 &  0.07 &   -   &  28.2$^{+  0.8}_{-  0.4}$ &  10.6 &  3.80 \\
    5 &   52.88047 &  -27.47975 &  1.70 &     0.0014 &   4.4 &   52.88068 &  -27.47972 & 11.841 &     0.0002 &   4.1 &   -   &   -   &             -             &   -   &   -   \\
    8 &   53.95480 &  -27.82182 & 11.00 &     0.0005 &   4.6 &   53.95509 &  -27.82199 &  9.762 &     0.0004 &   5.3 &   -   &   -   &             -             &   -   &   -   \\
    9 &      -     &      -     &   -   &     -      &   -   &   52.31905 &  -27.65615 & 13.678 & $<10^{-4}$ &   3.2 &  0.07 &   -   &  26.6$^{+  0.8}_{-  0.4}$ &  6.28 &  4.46 \\
   10 &      -     &      -     &   -   &     -      &   -   &   52.20959 &  -26.61418 & 19.408 & $<10^{-4}$ &   4.2 &   -   &   -   &             -             &   -   &   -   \\
   11 &   53.60244 &  -27.75840 &  0.40 &     0.0013 &   2.7 &   53.60240 &  -27.75882 &  4.057 &     0.0009 &   4.1 &   -   &   -   &             -             &   -   &   -   \\
\enddata
\end{deluxetable*}

\begin{deluxetable*}{cccccccccccccccc}
\tablenum{2}
\tabletypesize{\scriptsize}
\tablecaption{{\em continued} -- Radio and 24 $\mu$m counterparts 
to the $\geq 5 \sigma$ BLAST sources outside the FIDEL region}
\tablehead{
  \colhead{ID} &
  \colhead{$\alpha$\,(radio)} &
  \colhead{$\delta$\,(radio)} &
  \colhead{$f_{\rm r}$} &
  \colhead{$P_{\rm r}$} &
  \colhead{$d_{\rm r}$} &
  \colhead{$\alpha$\,(24 $\mu$m)} &
  \colhead{$\delta$\,(24 $\mu$m)} &
  \colhead{$f_{24}$} &
  \colhead{$P_{24}$} &
  \colhead{$d_{24}$} &
  \colhead{z$_{\rm RR}$} &
  \colhead{z$_{\rm 17}$} &
  \colhead{$T$} &
  \colhead{$L_{\rm FIR}$} &
  \colhead{$\chi^2$} \\
}
\startdata
   87 &      -     &      -     &   -   &     -      &   -   &   51.94289 &  -26.96452 &  4.102 &     0.0048 &  10.4 &   -   &   -   &             -             &   -   &   -   \\
   88 &      -     &      -     &   -   &     -      &   -   &   54.15564 &  -28.68730 & 11.775 &     0.0007 &   8.7 &   -   &   -   &             -             &   -   &   -   \\
   90 &   52.07532 &  -27.71906 &  1.50 &     0.0016 &   4.5 &   52.07560 &  -27.71935 &  5.607 &     0.0003 &   3.2 &   -   &   -   &             -             &   -   &   -   \\
   92 &   53.17420 &  -28.09792 &  0.30 &     0.0081 &   8.0 &   53.17420 &  -28.09762 &  0.827 &     0.0330 &   8.7 &  0.30 &   -   &  20.5$^{+  2.8}_{-  2.0}$ &  8.84 &  3.26 \\
   93 &      -     &      -     &   -   &     -      &   -   &   53.53340 &  -27.59049 &  0.830 &     0.0523 &  11.9 &   -   &   -   &             -             &   -   &   -   \\
   94 &   53.46999 &  -27.72898 &  0.20 &     0.0284 &  22.4 &   53.46997 &  -27.72978 &  1.702 &     0.0476 &  20.7 &   -   &   -   &             -             &   -   &   -   \\
   95 &      -     &      -     &   -   &     -      &   -   &   53.42970 &  -27.15331 &  6.743 &     0.0008 &   6.1 &   -   &   -   &             -             &   -   &   -   \\
   96 &   53.40470 &  -27.48562 &  0.30 &     0.0173 &  14.1 &   53.40502 &  -27.48516 &  3.293 &     0.0098 &  13.1 &   -   &   -   &             -             &   -   &   -   \\
\enddata
\end{deluxetable*}

\begin{deluxetable*}{cccccccccccccccc}
\tablenum{2}
\tabletypesize{\scriptsize}
\tablecaption{{\em continued} -- Radio and 24 $\mu$m counterparts 
to the $\geq 5 \sigma$ BLAST sources outside the FIDEL region}
\tablehead{
  \colhead{ID} &
  \colhead{$\alpha$\,(radio)} &
  \colhead{$\delta$\,(radio)} &
  \colhead{$f_{\rm r}$} &
  \colhead{$P_{\rm r}$} &
  \colhead{$d_{\rm r}$} &
  \colhead{$\alpha$\,(24 $\mu$m)} &
  \colhead{$\delta$\,(24 $\mu$m)} &
  \colhead{$f_{24}$} &
  \colhead{$P_{24}$} &
  \colhead{$d_{24}$} &
  \colhead{z$_{\rm RR}$} &
  \colhead{z$_{\rm 17}$} &
  \colhead{$T$} &
  \colhead{$L_{\rm FIR}$} &
  \colhead{$\chi^2$} \\
}
\startdata
  203 &   53.87595 &  -28.18568 &  0.80 &     0.0136 &  15.1 &   53.87598 &  -28.18580 &  0.376 &     0.1716 &  15.6 &   -   &   -   &             -             &   -   &   -   \\
  204 &   53.40178 &  -27.73207 &  0.22 &     0.0031 &   4.0 &   53.40199 &  -27.73190 &  0.480 &     0.0286 &   4.8 &   -   &  1.09 &  26.2$^{+  3.6}_{-  3.2}$ &  128. &  3.02 \\
  205 &      -     &      -     &   -   &     -      &   -   &   51.80349 &  -28.85086 &  1.303 &     0.0063 &   4.6 &   -   &   -   &             -             &   -   &   -   \\
  207 &   53.47462 &  -27.93015 &  0.30 &     0.0070 &   7.2 &   53.47474 &  -27.93006 &  1.199 &     0.0161 &   7.5 &   -   &   -   &             -             &   -   &   -   \\
  208 &      -     &      -     &   -   &     -      &   -   &   52.56557 &  -27.66277 &  0.533 &     0.0354 &   6.2 &   -   &   -   &             -             &   -   &   -   \\
  210 &   53.39663 &  -27.80555 &  0.06 &     0.0105 &   7.8 &   53.39699 &  -27.80564 &  0.645 &     0.0371 &   7.5 &   -   &  1.16 &  27.8$^{+  4.4}_{-  4.0}$ &  105. &  12.4 \\
  212 &      -     &      -     &   -   &     -      &   -   &   52.86584 &  -28.17471 &  0.932 &     0.0041 &   2.7 &  0.99 &   -   &  37.5$^{+  3.6}_{-  4.0}$ &  302. &  6.60 \\
  213 &   53.51498 &  -27.65649 &  0.20 &     0.0187 &  14.2 &   53.51506 &  -27.65668 &  0.526 &     0.1182 &  14.8 &   -   &   -   &             -             &   -   &   -   \\
\enddata
\end{deluxetable*}

\section{Multi-wavelength BLAST catalog}
\label{app_blast_cat}

Table \ref{tab_blast_cat} lists the 350 sources in the
multi-wavelength catalog of $\geq 5 \sigma$ BLAST sources in BGS-Wide
and BGS-Deep centered on the CDFS. See Section \ref{sec_blast_data}
for a description of how this catalog was created.  All sources are
detected with a significance of $\geq 5 \sigma$ in at least one of the
bands.

\begin{deluxetable*}{cccccccc}
\tabletypesize{\scriptsize}
\tablecaption{The multi-wavelength catalog of $\geq 5 \sigma$
BLAST sources in BGS-Wide and BGS-Deep. 
\label{tab_blast_cat} }
\tablecomments{
Reading from left to right, columns are: Internal ID as used throughout
this paper, official combined BLAST source ID then flux and $1 \sigma$ 
flux error for 250, 350 and 500 $\mu$m. To make the list, a source must
be detected at $\geq 5 \sigma$ in at least one band.
Note that the fluxes quoted
here have not been de-boosted and that errors are purely instrumental,
neglecting the effects of source confusion (see Section 
\ref{sec_sed_fitting} for further discussion).}
\tablehead{
  \colhead{ID} &
  \colhead{BLAST ID} &
  \colhead{$f_{250}$/Jy} &
  \colhead{$\sigma_{250}$/Jy} &
  \colhead{$f_{350}$/Jy} &
  \colhead{$\sigma_{350}$/Jy} &
  \colhead{$f_{500}$/Jy} &
  \colhead{$\sigma_{500}$/Jy} 
}
\startdata
  1 & BLAST J032921-280803 & 0.997 & 0.033 & 0.472 & 0.026 & 0.196 & 0.018\\
  2 & BLAST J032956-284631 & 1.056 & 0.038 & 0.536 & 0.029 & 0.227 & 0.021\\
  3 & BLAST J032741-282325 & 0.639 & 0.036 & 0.240 & 0.029 & 0.084 & 0.019\\
  4 & BLAST J033235-275530 & 0.177 & 0.011 & 0.075 & 0.009 & 0.034 & 0.006\\
  5 & BLAST J033131-272842 & 0.234 & 0.015 & 0.101 & 0.013 & 0.038 & 0.009\\
  6 & BLAST J033229-274415 & 0.157 & 0.011 & 0.079 & 0.008 & 0.043 & 0.006\\
  7 & BLAST J033250-273420 & 0.159 & 0.011 & 0.089 & 0.009 &   -   &   -  \\
  8 & BLAST J033548-274920 & 0.478 & 0.039 & 0.198 & 0.030 & 0.114 & 0.021\\
\enddata
\end{deluxetable*}

\begin{deluxetable*}{cccccccc}
\tablenum{3}
\tabletypesize{\scriptsize}
\tablecaption{{\em continued} -- The multi-wavelength catalog 
of $\geq 5 \sigma$ BLAST sources in BGS-Wide and BGS-Deep}
\tablehead{
  \colhead{ID} &
  \colhead{BLAST ID} &
  \colhead{$f_{250}$/Jy} &
  \colhead{$\sigma_{250}$/Jy} &
  \colhead{$f_{350}$/Jy} &
  \colhead{$\sigma_{350}$/Jy} &
  \colhead{$f_{500}$/Jy} &
  \colhead{$\sigma_{500}$/Jy} 
}
\startdata
  69 & BLAST J033153-281036 & 0.078 & 0.012 &   -   &   -   &   -   &   -  \\
  70 & BLAST J033111-284835 & 0.252 & 0.039 & 0.093 & 0.029 &   -   &   -  \\
  71 & BLAST J033140-272937 & 0.086 & 0.013 & 0.074 & 0.011 &   -   &   -  \\
  72 & BLAST J033120-273344 & 0.083 & 0.013 & 0.043 & 0.011 &   -   &   -  \\
  73 & BLAST J033158-273519 & 0.071 & 0.011 & 0.058 & 0.009 & 0.057 & 0.006\\
  74 & BLAST J033237-273531 & 0.071 & 0.011 & 0.058 & 0.009 & 0.046 & 0.006\\
  75 & BLAST J033115-273905 & 0.076 & 0.012 &   -   &   -   &   -   &   -  \\
  76 & BLAST J033328-273949 & 0.072 & 0.011 & 0.033 & 0.009 & 0.025 & 0.006\\
\enddata
\end{deluxetable*}

\begin{deluxetable*}{cccccccc}
\tablenum{3}
\tabletypesize{\scriptsize}
\tablecaption{{\em continued} -- The multi-wavelength catalog 
of $\geq 5 \sigma$ BLAST sources in BGS-Wide and BGS-Deep}
\tablehead{
  \colhead{ID} &
  \colhead{BLAST ID} &
  \colhead{$f_{250}$/Jy} &
  \colhead{$\sigma_{250}$/Jy} &
  \colhead{$f_{350}$/Jy} &
  \colhead{$\sigma_{350}$/Jy} &
  \colhead{$f_{500}$/Jy} &
  \colhead{$\sigma_{500}$/Jy} 
}
\startdata
  145 & BLAST J033211-275859 & 0.060 & 0.011 &   -   &   -   &   -   &   -  \\
  146 & BLAST J033000-275347 & 0.153 & 0.028 &   -   &   -   &   -   &   -  \\
  147 & BLAST J033110-274302 & 0.063 & 0.012 & 0.052 & 0.009 &   -   &   -  \\
  148 & BLAST J033104-275001 & 0.062 & 0.011 & 0.045 & 0.009 &   -   &   -  \\
  149 & BLAST J033612-281046 & 0.206 & 0.038 & 0.107 & 0.030 & 0.083 & 0.021\\
  150 & BLAST J033416-280216 & 0.081 & 0.015 & 0.043 & 0.011 &   -   &   -  \\
  151 & BLAST J033109-280045 & 0.066 & 0.012 &   -   &   -   & 0.036 & 0.007\\
  152 & BLAST J033648-271936 & 0.205 & 0.038 & 0.116 & 0.029 & 0.064 & 0.021\\
\enddata
\end{deluxetable*}

\begin{deluxetable*}{cccccccc}
\tablenum{3}
\tabletypesize{\scriptsize}
\tablecaption{{\em continued} -- The multi-wavelength catalog 
of $\geq 5 \sigma$ BLAST sources in BGS-Wide and BGS-Deep}
\tablehead{
  \colhead{ID} &
  \colhead{BLAST ID} &
  \colhead{$f_{250}$/Jy} &
  \colhead{$\sigma_{250}$/Jy} &
  \colhead{$f_{350}$/Jy} &
  \colhead{$\sigma_{350}$/Jy} &
  \colhead{$f_{500}$/Jy} &
  \colhead{$\sigma_{500}$/Jy} 
}
\startdata
  222 & BLAST J032753-284023 &   -   &   -   & 0.168 & 0.029 &   -   &   -  \\
  223 & BLAST J033423-274409 & 0.064 & 0.014 & 0.060 & 0.010 &   -   &   -  \\
  224 & BLAST J033134-274628 &   -   &   -   & 0.049 & 0.009 & 0.021 & 0.006\\
  225 & BLAST J033123-275233 &   -   &   -   & 0.048 & 0.008 &   -   &   -  \\
  226 & BLAST J033723-274021 &   -   &   -   & 0.163 & 0.029 &   -   &   -  \\
  227 & BLAST J033134-280059 & 0.038 & 0.011 & 0.048 & 0.009 &   -   &   -  \\
  228 & BLAST J033635-283558 & 0.126 & 0.036 & 0.154 & 0.028 & 0.104 & 0.019\\
  229 & BLAST J033102-274456 & 0.059 & 0.012 & 0.055 & 0.010 &   -   &   -  \\
\enddata
\end{deluxetable*}

\begin{deluxetable*}{cccccccc}
\tablenum{3}
\tabletypesize{\scriptsize}
\tablecaption{{\em continued} -- The multi-wavelength catalog 
of $\geq 5 \sigma$ BLAST sources in BGS-Wide and BGS-Deep}
\tablehead{
  \colhead{ID} &
  \colhead{BLAST ID} &
  \colhead{$f_{250}$/Jy} &
  \colhead{$\sigma_{250}$/Jy} &
  \colhead{$f_{350}$/Jy} &
  \colhead{$\sigma_{350}$/Jy} &
  \colhead{$f_{500}$/Jy} &
  \colhead{$\sigma_{500}$/Jy} 
}
\startdata
  297 & BLAST J033028-275625 & 0.084 & 0.019 & 0.071 & 0.015 & 0.065 & 0.010\\
  299 & BLAST J033336-275751 & 0.050 & 0.011 & 0.038 & 0.009 & 0.038 & 0.006\\
  300 & BLAST J033114-273556 &   -   &   -   &   -   &   -   & 0.046 & 0.007\\
  301 & BLAST J033128-273745 &   -   &   -   & 0.038 & 0.009 & 0.040 & 0.006\\
  302 & BLAST J033552-275511 &   -   &   -   &   -   &   -   & 0.130 & 0.021\\
  303 & BLAST J033121-275803 &   -   &   -   &   -   &   -   & 0.038 & 0.006\\
  304 & BLAST J033231-280437 & 0.049 & 0.011 & 0.037 & 0.009 & 0.037 & 0.006\\
  306 & BLAST J033154-284219 &   -   &   -   &   -   &   -   & 0.118 & 0.019\\
\enddata
\end{deluxetable*}

\begin{flushleft}
{\bf Acknowledgments}
\end{flushleft}

We acknowledge the support of NASA through grant numbers NAG5-12785,
NAG5- 13301, and NNGO-6GI11G, the NSF Office of Polar Programs, the
Canadian Space Agency and the Natural Sciences and Engineering
Research Council (NSERC) of Canada.  This work is based in part on
observations made with the Spitzer Space Telescope, which is operated
by the Jet Propulsion Laboratory, California Institute of Technology
under a contract with NASA. SD is supported by the UK Science and
Technology Facilities Council (STFC).  We thank Guilane Lagache for
advice on implementation of her galaxy evolution models.


{}


\end{document}